\documentclass[%
reprint,
amsmath,amssymb,
aps,
prl,
]{revtex4-1}

\usepackage{xcolor}
\usepackage{graphicx}
\usepackage{dcolumn}
\usepackage{bm}

\begin{document}
	
\title{Kibble-Zurek behavior in disordered Chern insulators}

\author{ Lara Ul\v{c}akar}

\affiliation{Jozef Stefan Institute, Jamova 39, SI-1000 Ljubljana, Slovenia}

\affiliation{Faculty for Mathematics and Physics, University of Ljubljana, Jadranska
	19, SI-1000 Ljubljana, Slovenia}

\email{lara.ulcakar@ijs.si}

\author{Jernej Mravlje}

\affiliation{Jozef Stefan Institute, Jamova 39, SI-1000 Ljubljana, Slovenia}

\author{Toma\v{z} Rejec}

\affiliation{Jozef Stefan Institute, Jamova 39, SI-1000 Ljubljana$ $, Slovenia}

\affiliation{Faculty for Mathematics and Physics, University of Ljubljana, Jadranska
	19, SI-1000 Ljubljana, Slovenia}

\date{\today}	

\begin{abstract}
Even though no local order parameter in the sense of the Landau theory
exists for topological quantum phase transitions in Chern insulators, the highly non-local Berry curvature
exhibits critical behavior near a quantum critical point. We investigate the critical properties of its real space analog, the local Chern marker, in weakly disordered Chern insulators. Due to disorder, inhomogeneities appear in the spatial distribution of the local Chern marker. Their size exhibits power-law scaling with the critical exponent matching the one extracted from the Berry curvature of a clean system. We drive the system slowly through such a quantum phase transition. The characteristic size of inhomogeneities in the non-equilibrium post-quench state obeys the Kibble-Zurek scaling. In this setting, the local Chern marker thus does behave in a similar way as a local order parameter for a symmetry breaking second order phase transition. The Kibble-Zurek scaling also holds for the inhomogeneities in the spatial distribution of excitations and of the orbital polarization.

\end{abstract}

\maketitle

The discovery of topological insulators \cite{HaldaneModel,SHallRashba,KaneMele} has sparked great interest due to their novel properties that could furthermore be used for practical applications \cite{MemCell,MagSwitch,Thermoel,ElInterconn,AppRev}. Topological systems have been realized in solid state  \cite{Novoselov07,HgTeExp,Yu10,InAsExp,Chang13,Bismuthene,Tokura19} and in cold atoms \cite{Tarruell12,Aidelsburger13,Miyake13,Goldman13,Dauphin13,Jotzu14,Wu16,Flaschner16}.
Recently, a lot of attention has been given to 
dynamical critical properties after quenches across topological phase transitions in topological insulators \cite{Caio2015,Mitra16,Caio2016,Unal16,zoller,Privitera16,Wilson16,Wang16,WangNJP16,Duta17,Schuler17,Ulcakar2018,McGinleyPRL18,McGinley18,Liou18,Ulcakar19}, superconductors \cite{Bermudez10,DeGottardi11,Sacramento14} and $p+ip$ superfluids \cite{PxPySuperFluid,Foster14}. An important observation of interest to this paper was made in  Refs.~\cite{Damski05,Dutta10,Ulcakar2018,Ulcakar19,Marincek19} which showed that the number of excitations after a slow quench follows the Kibble-Zurek (KZ) scaling.

Developed by Kibble \cite{Kibble} as a cosmological theory describing the formation of the early universe and applied to condensed matter systems by Zurek \cite{Zurek,Zurek05}, the KZ mechanism describes non-equilibrium properties of a system that was driven in a finite time $\tau$ over a symmetry breaking second order phase transition. In equilibrium the relaxation time $\tau_r$ and the correlation length $\xi$ diverge as a function of the control parameter $u$ approaching the critical point $u_c$ by a power determined by critical exponents $z$ and $\nu$, $\tau_r\sim|u-u_c|^{-z\nu}$ and $\xi\sim|u-u_c|^{-\nu}$  \cite{Polkovnikov05}. Because of the divergence the system evolves nonadiabatically across the critical point. In the case of phase transitions with spontaneous symmetry breaking that entail a degeneracy of the ground state, such a process produces regions corresponding to different choices of the ground state. 
The size of the regions is set by the equilibrium correlation length $\xi(t_F)\sim\tau^{\nu/(1+z\nu)}$ at a "freeze-out" time $t_F\sim\tau^{z\nu/(1+z\nu)}$, an approximate time at which the system stopped evolving adiabatically. 
The KZ scaling was observed experimentally in tunnel Josephson junctions \cite{Monaco02,Monaco06}, multiferroics \cite{Chae12,Griffin12,Meier17}, ion Coulomb crystals \cite{Ulm13,Pyka13}, Bose-Einstein condensates \cite{Lamporesi13}, and in a Rydberg atom quantum simulator \cite{Keesling19}. 

\begin{figure*}
	\includegraphics[width=0.195\textwidth]{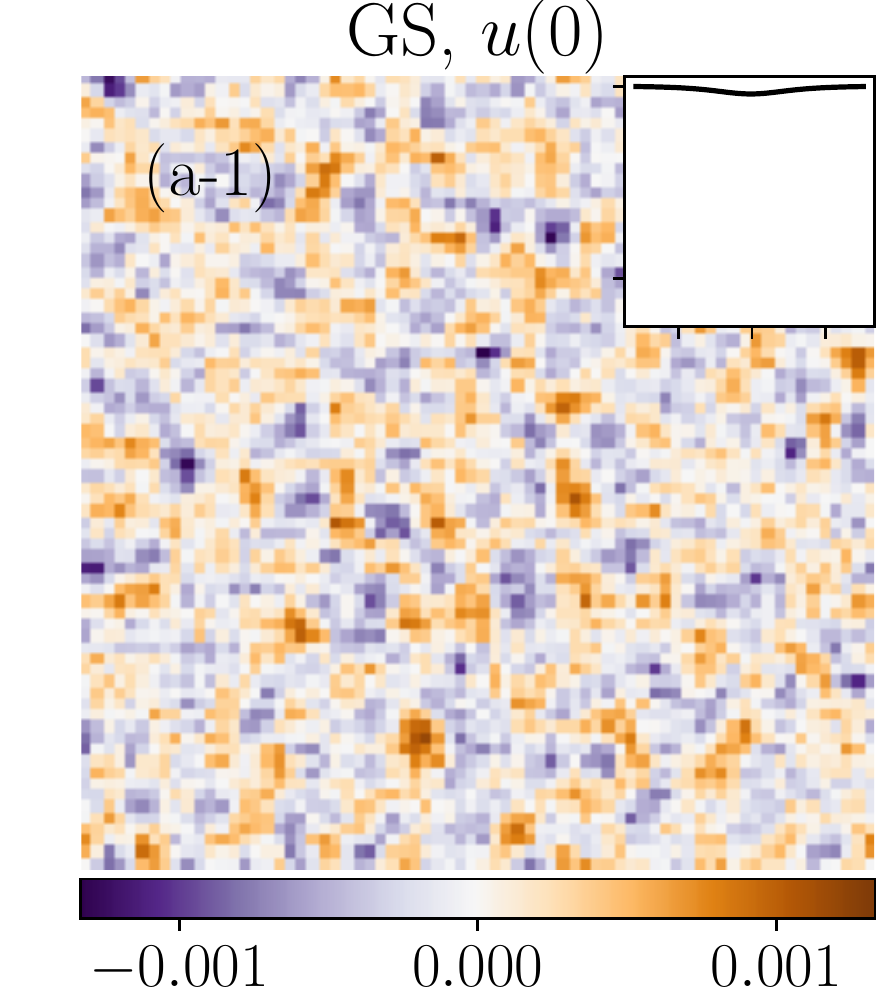}
\includegraphics[width=0.195\textwidth]{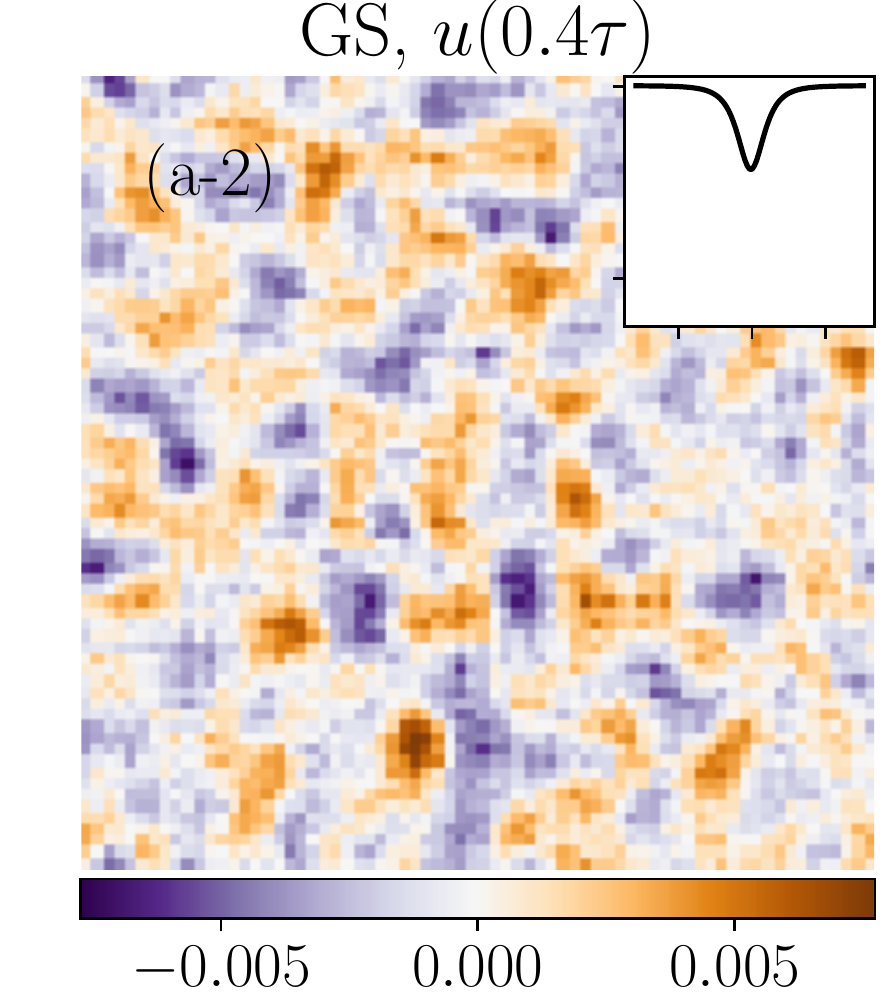}
\includegraphics[width=0.195\textwidth]{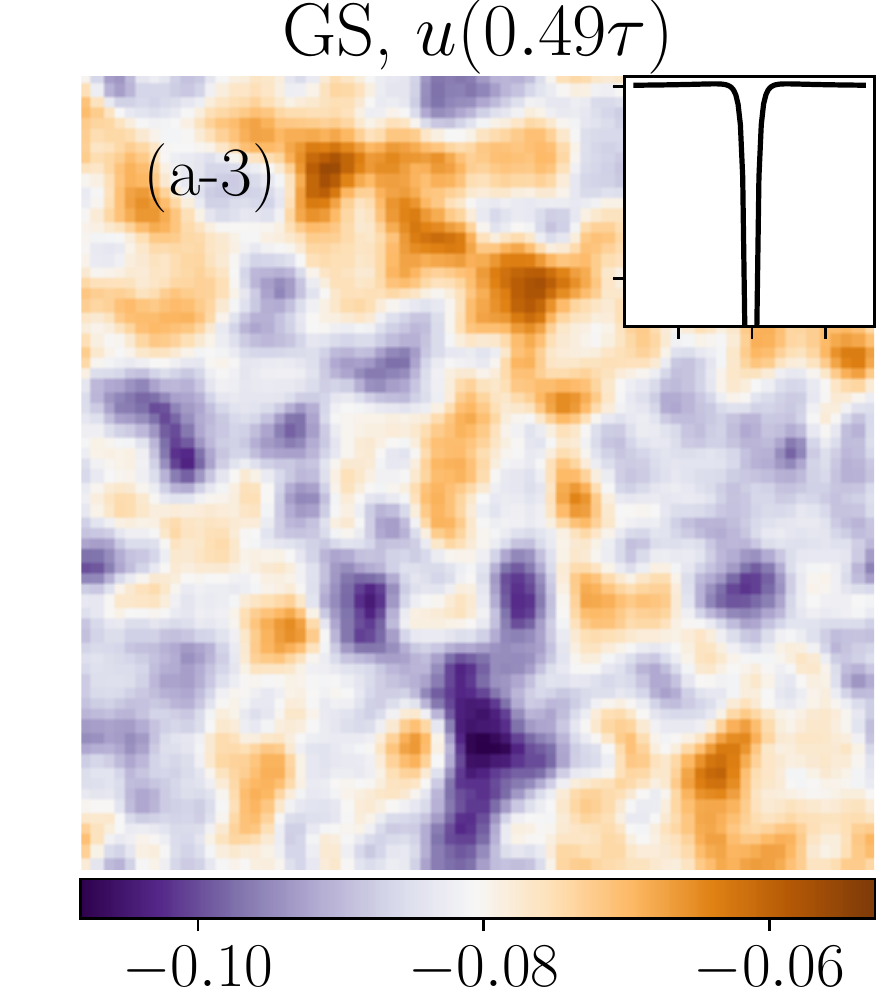}
\includegraphics[width=0.195\textwidth]{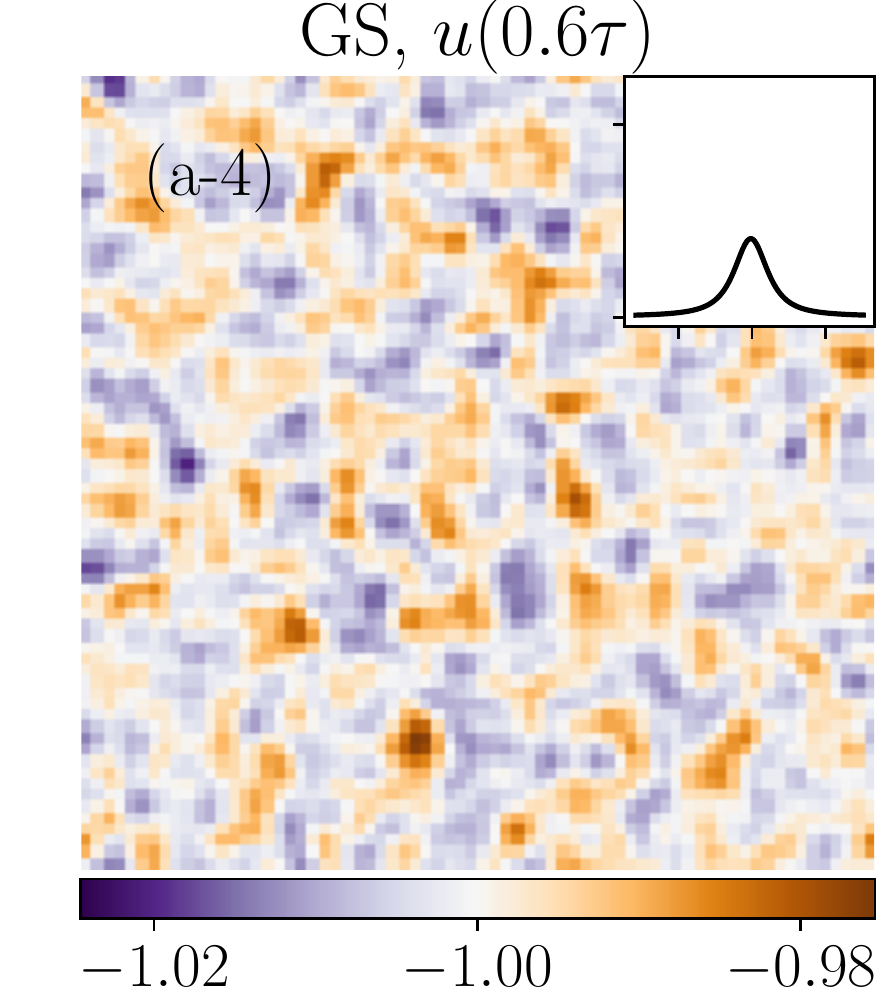}
\includegraphics[width=0.195\textwidth]{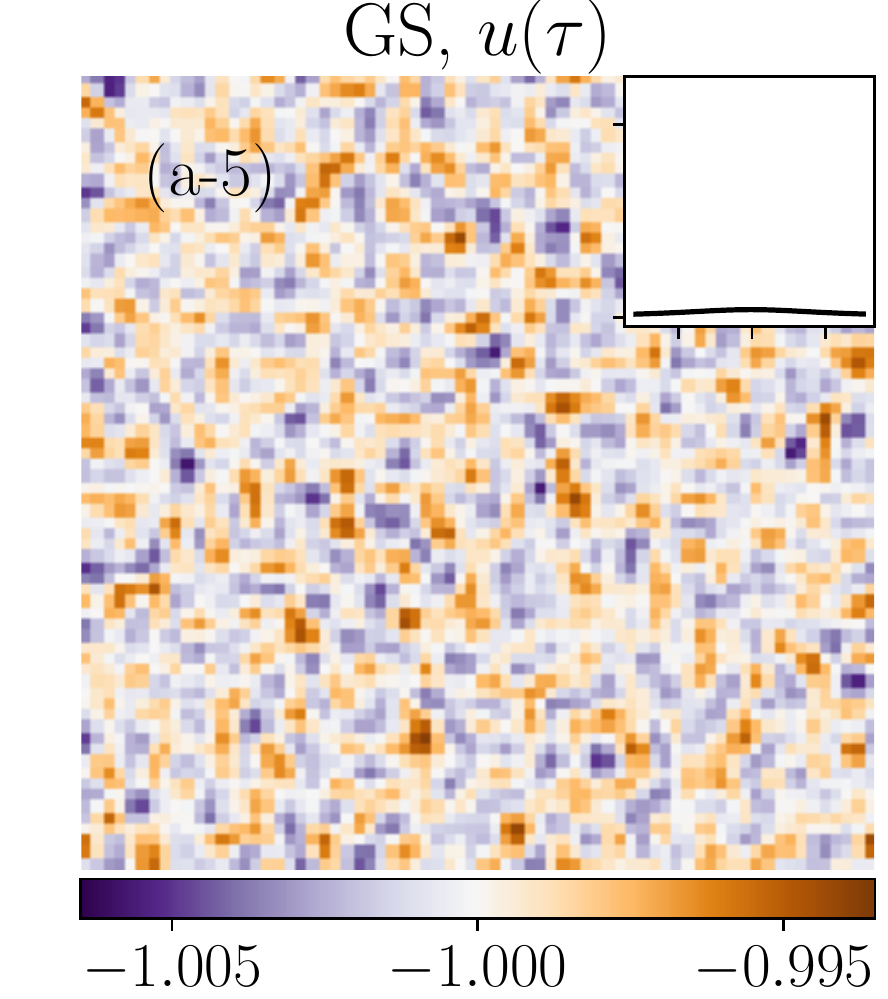}
\includegraphics[width=0.195\textwidth]{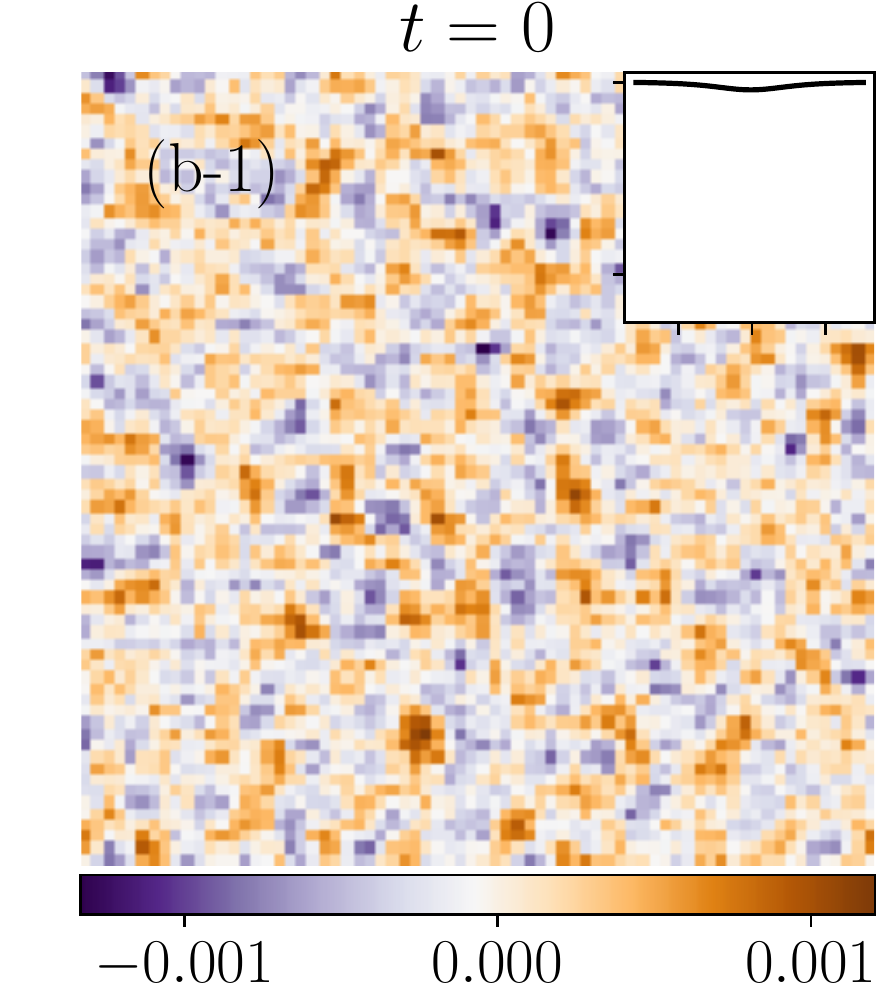}
\includegraphics[width=0.195\textwidth]{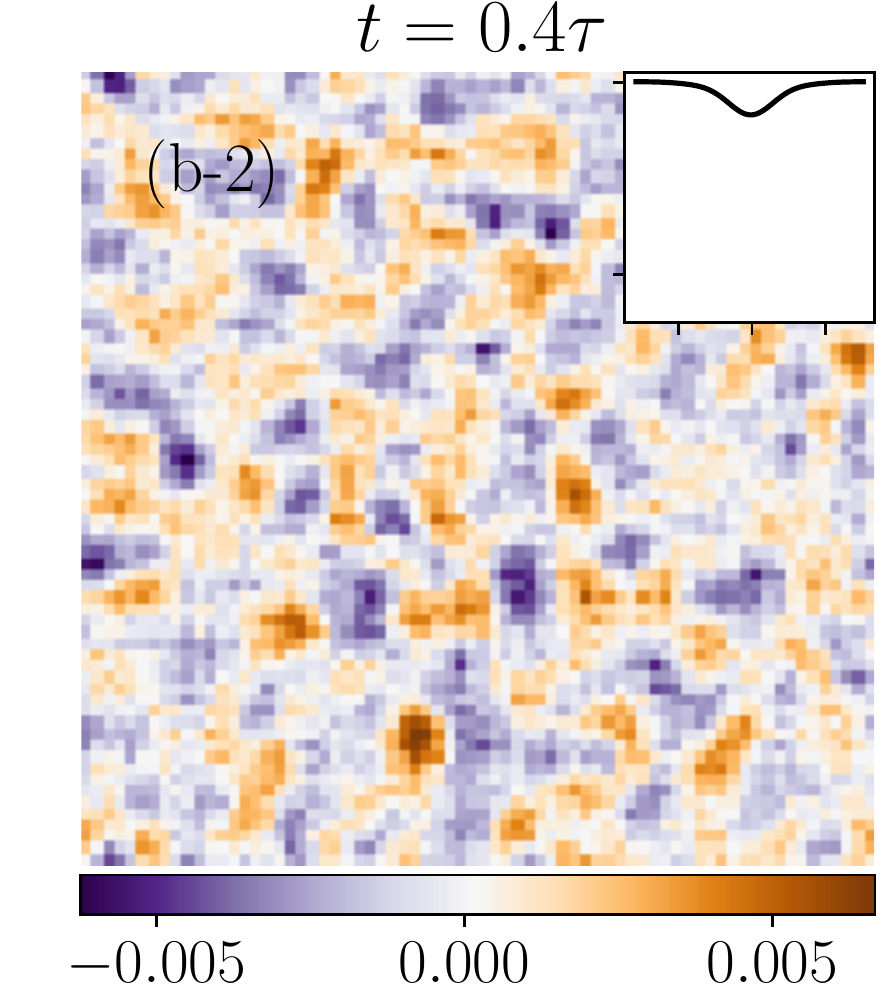}
\includegraphics[width=0.195\textwidth]{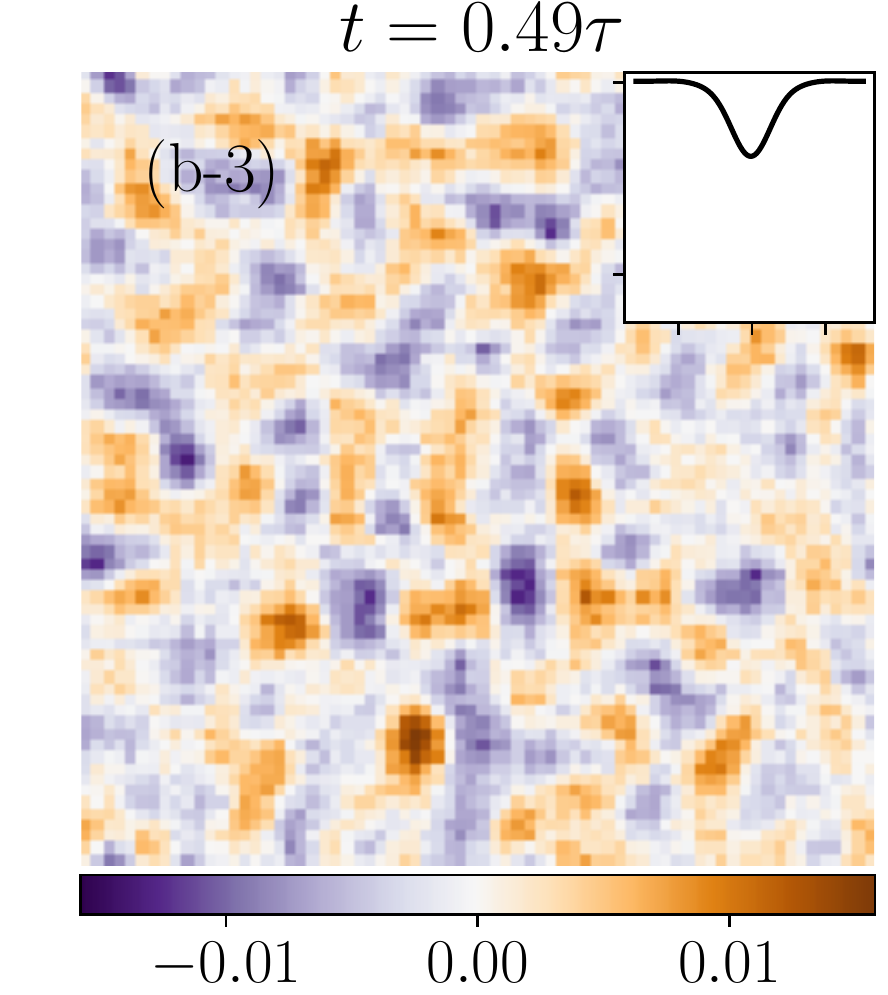}
\includegraphics[width=0.195\textwidth]{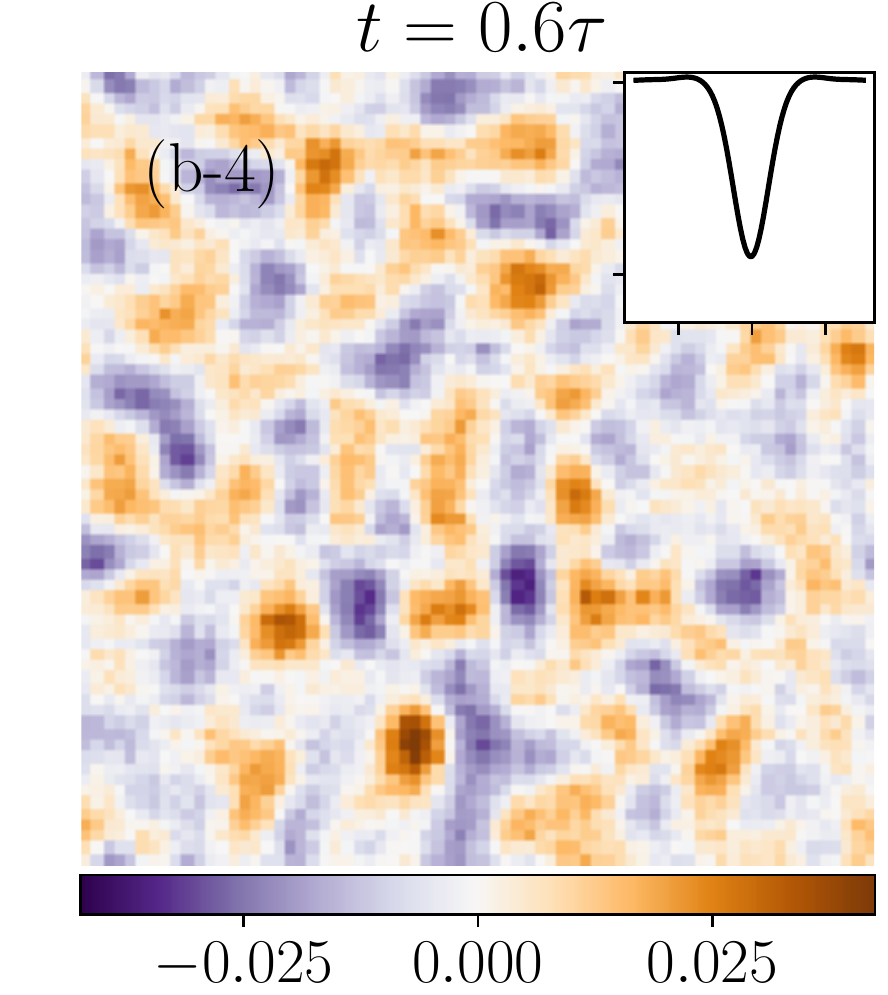}
\includegraphics[width=0.195\textwidth]{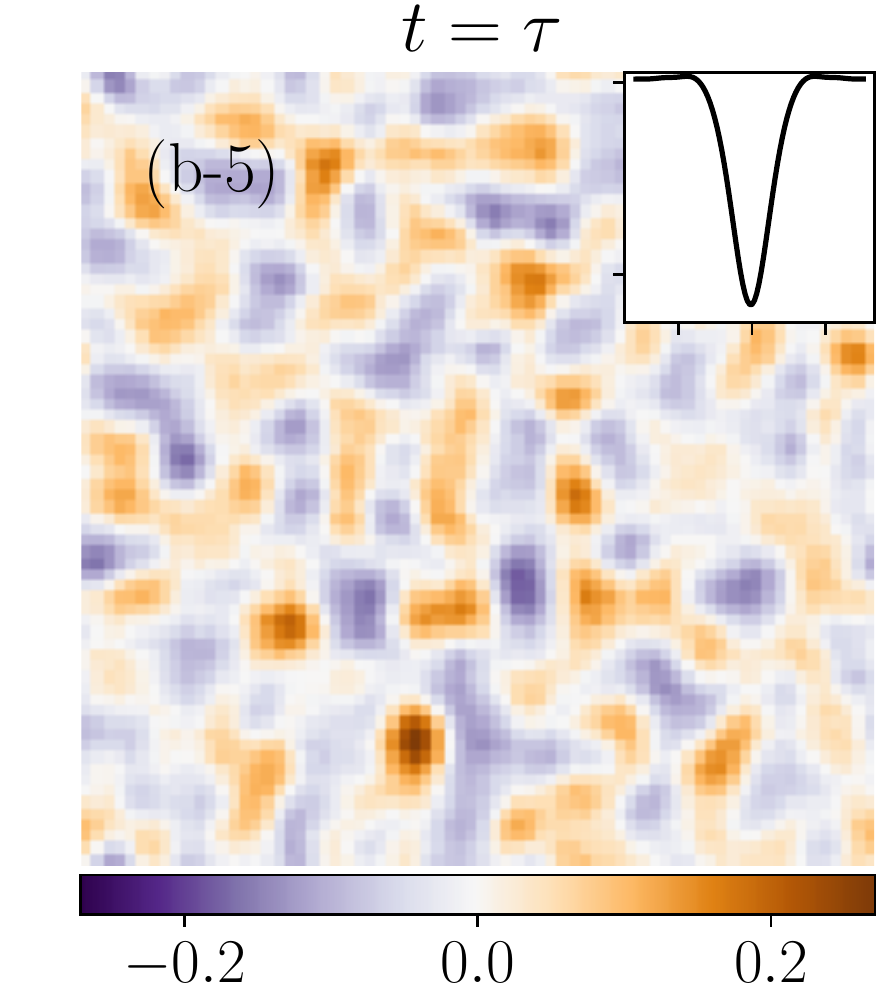}
	
	\caption{LCM (a) in the ground state of an instantaneous Hamiltonian and (b) in a non-equilibrium state during the quench with $\tau=20$: (1) at the beginning of the quench, (2) on the entry to the freeze-out zone, (3) near the critical point, (4) on the exit from the freeze-out zone, and (5) at the end of the quench. The system size is $N=70$ and the disorder amplitude is $\delta u_0=0.05$.
	The insets show the Berry curvature of corresponding states of a clean system along the $k_x=0$ line from $k_y=-\frac{\pi}{4}$ to $k_y=\frac{\pi}{4}.$ The range of the vertical axis is from zero to 0.005$\pi$ in panels (a-4) and (a-5), and from -0.005$\pi$ to zero in other panels.}
	
	\label{fig:movie}
\end{figure*}

Returning to topological insulators, the fact that the KZ scaling occurs in these systems is not expected on the first sight. Topological insulators have no spontaneously broken symmetry. They lack degeneracy of the ground state and have no local (Landau) order parameter. Instead, their phase is described by a quantized non-local order parameter, the topological invariant. Is it nevertheless possible to relate the KZ scaling to the freeze-out behavior and what are its manifestations in the real space?
In Chern insulators, for example, the critical increase of a length scale has been noticed in the Berry curvature and in the overlap of Wannier states \cite{Chen16,Chen17,Chen19}.
Some hope for establishing the analogy with systems with a local order parameter stems also from the discovery of the local Chern marker (LCM) that was introduced in Ref.~\cite{Cmarker} as a local indicator of the topological phase in Chern insulators. Furthermore, in Ref.~\cite{Caio19} Caio {\it et al.} showed that in equilibrium, the LCM exhibits a length scale that grows as $|u-u_c|^{-\nu}$ close to a topological phase transitions. An open question is, how does the LCM behave during a slow quench?
 
In this paper we address the KZ mechanism in a Chern insulator. To reveal the KZ physics directly in real space, we calculate the LCM in the presence of weak disorder which breaks the translational symmetry. The disorder leads to the appearance of inhomogeneities in the LCM. We show that in the ground state these exhibit a length scale shown in Fig.~\ref{fig:movie}(a) that grows as $|u-u_c|^{-1}$ as the topological transition is approached. Then we study a quench where we drive the system across a critical point in a finite time $\tau$. We evaluate the LCM during the quench and find inhomogeneities [Fig.~\ref{fig:movie}(b)], the size of which in the post-quench state scales as $\tau^{1/2}$, as predicted by the KZ mechanism. The growth of inhomogeneities during the quench exhibits the freeze-out behavior, which establishes an almost full analogy to the dynamical critical behavior of systems with a spontaneously broken symmetry. The same behavior is found also in the clean system where the length scale is extracted from the Berry curvature evaluated during the quench.

\textit{Ground state.}---We first examine the ground-state critical behavior in the vicinity of a topological quantum phase transition in a Chern insulator described by the disordered Qi-Wu-Zhang (QWZ) model \cite{QWZmodel} 
\begin{equation}
\begin{split}
\hat H=&\sum_{\mathbf r}|x,y\rangle\langle x,y|\otimes[u+\delta u(\mathbf r)]\hat\sigma_z+\\
+&\sum_{\mathbf r}\left(|x+1,y\rangle\langle x,y|\otimes\tfrac{\hat\sigma_z+i\hat\sigma_x}{2}+\textrm{h.c.}\right)+\\
+&\sum_{\mathbf r}\left(|x,y+1\rangle\langle x,y|\otimes\tfrac{\hat \sigma_z+i\hat \sigma_y}{2}+\textrm{h.c.}\right)
\label{eq:QWZreal}
\end{split}
\end{equation}
where $\mathbf{r}=(x,y)$ are the Bravais lattice vectors of a square lattice, measured in units of the lattice constant. The system size is $N\times N$ unit cells and periodic boundary conditions are assumed. Each unit cell hosts two orbitals $|\mathbf{r},\sigma\rangle$, $\sigma\in\left\{A,B\right\}$, and the Pauli matrices $\hat\sigma_i$, $i\in\{x,y,z\}$, act on these orbital degrees of freedom. The disorder is present on the staggered orbital binding energies $u+\delta u(\mathbf r)$, where $\delta u(\mathbf r)$ are uncorrelated and uniformly distributed on the interval $[-\delta u_0,\delta u_0]$. The disorder is assumed to be weak enough that the Anderson localization length is much longer than the system size. In absence of disorder, the QWZ model was experimentally realized in ultracold atoms \cite{Liu14,Wu16}.

In a clean system, the QWZ Hamiltonian for a particular wavevector $\mathbf{k}=(k_x,k_y)$ from the first Brillouin zone (BZ) is
\begin{equation}
\hat{H}(\mathbf{k})=(u+\cos{k_x}+\cos{k_y})\hat{\sigma}_z+
\sin{k_x}\hat{\sigma}_x+\sin{k_y}\hat{\sigma}_y.\label{eq:Hbulk}
\end{equation}
Its eigenstates $|\psi_\alpha(\mathbf{k})\rangle$, 
where the band index $\alpha$ distinguishes the valence ($\alpha=v$) from the conduction ($\alpha=c$) band, can be used to calculate the Berry curvature \cite{Prodan10} of the valence band,
\begin{equation}
\Omega(\mathbf{k})=i\textrm{Tr}\{\hat P(\mathbf k)[\partial_{k_x}\hat P(\mathbf k),\partial_{k_y}\hat P(\mathbf k)]\}
\end{equation}
with $\hat P(\mathbf k)=|\psi_v(\mathbf k)\rangle\langle \psi_v(\mathbf k)|$. In this work, we concentrate on the topological quantum phase transition that takes place at $u_c=-2$ where the topological invariant --~the Chern number $C=-\frac{1}{2\pi}\int_\textrm{BZ}\mathrm{d}\mathbf{k}\Omega(\mathbf{k})$~-- changes its value from $C=0$ at $u<u_c$ to $C=-1$ at $u_c<u<0$. 

The Chern number can also be calculated from the real-space analogue of the Berry curvature -- the local Chern marker \cite{Cmarker,Prodan10}
\begin{equation}
c(\mathbf{r})=2\pi i \sum_{\sigma}\langle\mathbf{r},\sigma|\hat{P}[-i[\hat{x},\hat{P}],-i[\hat{y},\hat{P}]]|\mathbf{r},\sigma\rangle
\label{LCM}
\end{equation}                
as $C=\lim_{N\to\infty}\frac{1}{N^2}\sum_\mathbf{r}c(\mathbf{r})$. Here $\hat x$ and $\hat y$ are the position operators and $\hat{P}=\sum_{n\in v}|\Psi_n\rangle\langle\Psi_n|$ is the projector onto the subspace spanned by eigenstates $|\Psi_n\rangle$ of the valence band. In clean systems in the thermodynamic limit the LCM is uniform and equals the Chern number \cite{Cmarker}. 

We calculated the LCM in the presence of a weak disorder $\delta u_0=0.05$ (for computational details see Supplemental Material S1 \cite{SM}). The real-space profiles of the LCM are shown in Fig.~\ref{fig:movie}(a) for several values of $u$, ranging from deep in the trivial phase, across the critical point (note that the weak disorder does not significantly move the critical point \cite{Yamakage11}) to deep in the topological phase. The profiles are inhomogeneous and feature regions where the LCM deviates above (brown) and below (blue) the clean system value. While the amplitude of those deviations is proportional to  $\delta u_0$, their size does not depend on the disorder strength (see Supplemental Material S2 \cite{SM}). The basic point is that, as the critical point is approached, the size of those regions grows. [In the topological phase the deviations of the LCM from its clean system value are dominated by a contribution proportional to disorder. In Fig.~\ref{fig:movie}(a-4) and Fig.~\ref{fig:movie}(a-5),  the disorder contribution is filtered out. Raw data is shown in Supplemental Material S2 \cite{SM}.]  

We measure the size of inhomogeneities $\xi_r$ by finding the distance where the disorder-averaged autocovariance function of a LCM profile drops below zero (see Supplemental Material S3 \cite{SM}). Fig.~\ref{fig:domGS} shows that $\xi_r$ exhibits a power-law scaling as $u$ approaches the critical point. Increasing the system size, the estimate of the scaling exponent approaches one. One can evaluate the correlation length also for a clean system by calculating the Berry curvature, which exhibits a peak at $\mathbf k=0$  [insets to Fig.~\ref{fig:movie}(a)]. The width of the peak $\xi_k^{-1}$ shrinks on approaching the critical point with the correlation length exponent $\nu=1$ \cite{Chen16,Ulcakar2018,nune1}, thus in agreement with what we find from the analysis of the inhomogeneities in the LCM. Analogous results are obtained by measuring the radius of the peak in the LCM profile around a single impurity (see Supplemental material S4 \cite{SM}). All this confirms that, in the presence of disorder, the criticality of the correlation length in Chern insulators is observable in real space via the LCM. 

\begin{figure}[h]
	\includegraphics[width=\columnwidth]{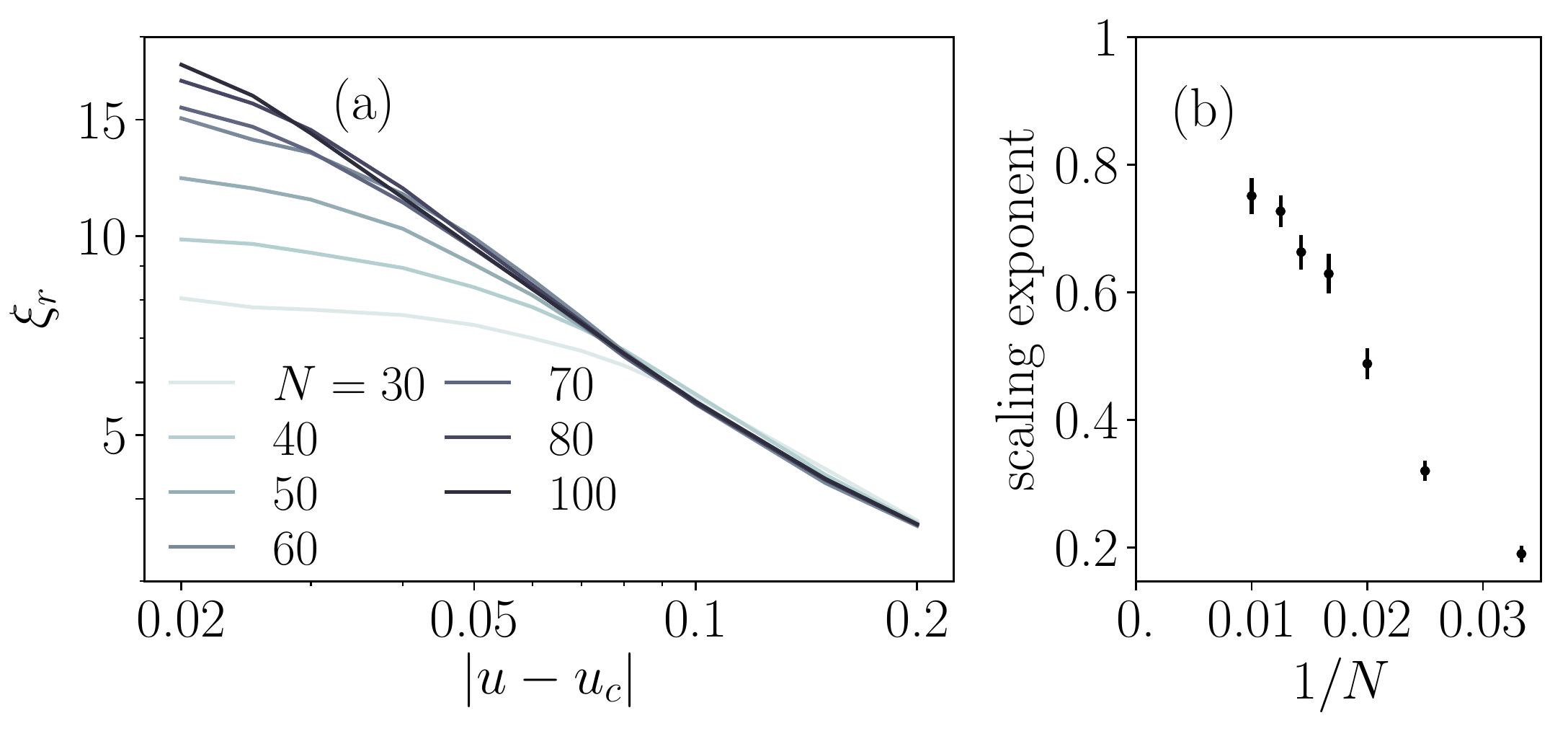}
	\caption{(a) $\xi_r$ in the ground state of the trivial phase, plotted for different $N$ in the log-log scale as a function of $|u-u_c|$. The average is performed over 100 disorder realizations with $\delta u_0=0.05$. (b) The scaling exponent, estimated by fitting the power-law to the $\xi_r$ vs. $|u-u_c|$ data in the interval $0.02\leq |u-u_c|\leq0.1$ and plotted as a function of $1/N$. Error bars are standard errors obtained by bootstrap resampling.}
	\label{fig:domGS}
\end{figure} 

\textit{Quenches.}---We perform quenches in systems initially in the ground state, i.e., with the valence band filled and the conduction band empty, starting in the trivial regime at $u_0=-2.5$, smoothly varying the parameter $u$ with time as $u(t)=u_0+(u_1-u_0) \sin^2(\frac{\pi}{2}\frac{t}{\tau})$, and ending up at $t=\tau$ in the topological regime at $u_1=-1.5$. 
We separately show that the length scale, predicted by the KZ mechanism, arises both in clean and in disordered systems. 

\begin{figure}[h]
	\includegraphics[width=123pt]{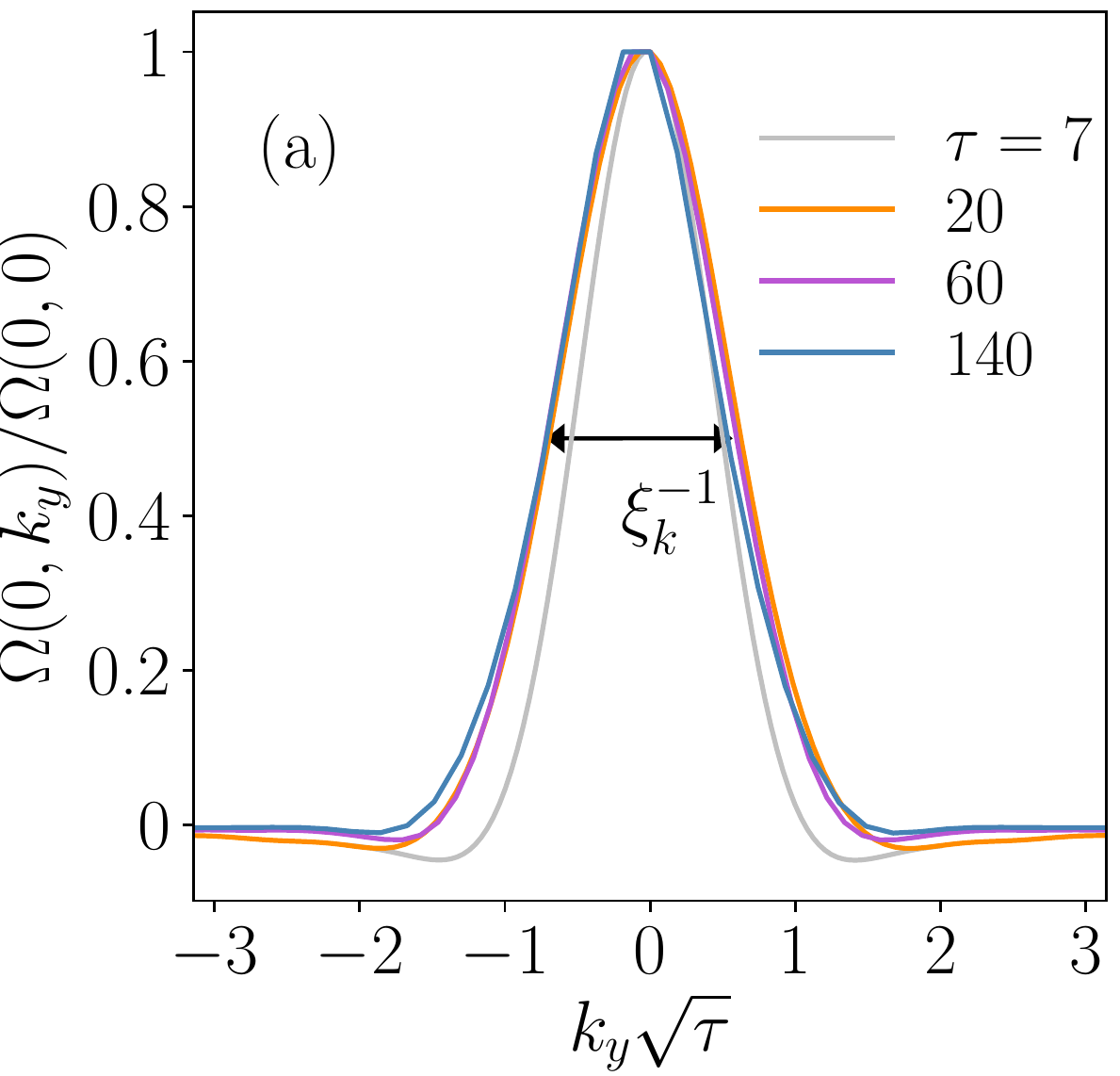}\includegraphics[width=125pt]{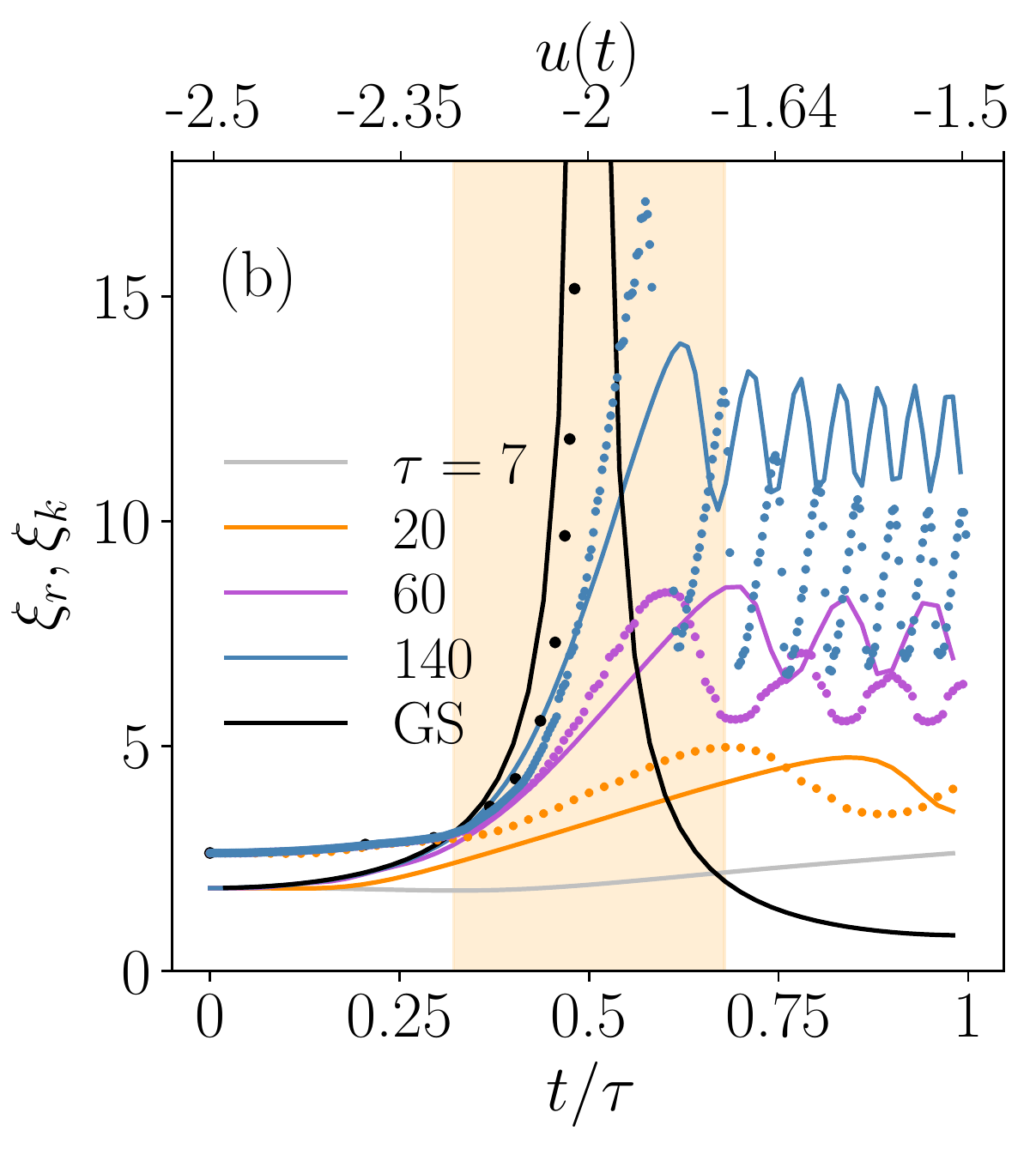}
	\caption{(a) Berry curvature along the $k_x=0$ line at the end of a quench for various $\tau$, plotted as a function of $k_y\sqrt{\tau}$. (b) Full lines show $\xi_k$ during quenches with various $\tau$ (colored) and $\xi_k$ of instantaneous ground states (black). Colored and black dots show $\xi_r$ of the LCM of a disordered system with $\delta u_0=0.05$ during quenches ($N=70$) and in instantaneous ground states ($N=100$, for $t/\tau < 0.5$), respectively. The shaded region is the freeze-out zone for the $\tau=20$ quench.} 
	\label{fig:berry}
\end{figure}

\textit{Quenches in clean systems.}--- We extract $\xi_k$ after the quench from the Berry curvature of non-equilibrium post-quench states, calculated by replacing the pre-quench eigenstates $|\psi_v(\mathbf k)\rangle$ in Eq.~(\ref{fig:berry}) with the corresponding time-evolved states.
The post-quench form of the Berry curvature for different quench times is shown in Fig.~\ref{fig:berry}(a). It exhibits a peak at $\mathbf k=0$ of the width proportional to $\tau^{-1/2}$, giving rise to a length scale proportional to $\tau^{1/2}$. Taking into account that  $z\nu=1$ \cite{Ulcakar2018}, this result follows the KZ scaling. The KZ mechanism relates the characteristic length with the density of point defects $n\propto \xi^{-2}(t_F)$, making our results in agreement with earlier work \cite{Damski05,Dutta10,Ulcakar2018,Ulcakar19}, which showed that excitation density is  $n_\mathrm{exc}=\frac{|u_1-u_0|}{8\pi\tau}$.


It is interesting to look at how the Berry curvature evolves during the quench  [insets to Fig.~\ref{fig:movie}(b)]. Fig.~\ref{fig:berry}(b) shows the evolution of $\xi_k$ of non-equilibrium states corresponding to different quench times (colored lines). This is compared to the critical behavior of $\xi_k$ of the corresponding instantaneous ground states (black line). Before entering the freeze-out zone (the shaded region for $\tau=20$), the system evolves adiabatically and its $\xi_k$ is equal to the ground-state one. In the freeze-out zone, the system stops evolving adiabatically and its $\xi_k$ starts to deviate from the ground-state one: it increases linearly with a speed of the order of the Fermi velocity, independent of the quench time. Only after the system exits the freeze-out zone, $\xi_k$ settles at an approximately constant value. 
Similar results were observed in quenches performed in a Rydberg atom quantum simulator \cite{Keesling19}. This behavior goes beyond what is known in the literature as the ``adiabatic-impulse approximation'' which identifies the saturation of the length scale with the entry to the freeze-out zone \cite{Zurek05}.

\textit{Quenches in disordered systems.}---The KZ length scale observed in a clean system manifests itself in a disordered system as the size of inhomogeneities in the post-quench LCM profile. 

Similarly to the Berry curvature, the LCM profiles can also be calculated during a quench by replacing in Eq.~(\ref{LCM}) the projector $\hat{P}$ onto the valence band with the projector onto the occupied subspace i.e,  pre-quench eigenstates $|\Psi_n \rangle$ are replaced with the corresponding time-evolved states.

The evolution of the LCM profile during the quench with $\tau=20$ is shown in Fig.~\ref{fig:movie}(b). Due to the conservation of the Chern number \cite{Rigol2015,Caio2015,Ulcakar2018, McGinley18}, the clean system value of the LCM is zero throughout the quench. Let us now focus on deviations from this value. Prior to entering the freeze-out zone, the system evolves adiabatically and the LCM profiles match the ground-state ones, shown in Fig.~\ref{fig:movie}(a). The growth of $\xi_r$ in the freeze-out zone, although present, lags behind that in instantaneous ground states. After the exit from the freeze-out zone, the profiles do not significantly change anymore. [The amplitude of the deviations of the LCM, however, grows strongly throughout the quench.] In Fig.~\ref{fig:berry}(b) $\xi_r$ during quenches with different $\tau$ and in the corresponding ground states are shown with colored and black dots, respectively. They are seen to follow roughly the behavior of $\xi_k$ (colored and black lines). 

\begin{figure}[t]
	\includegraphics[width=\columnwidth]{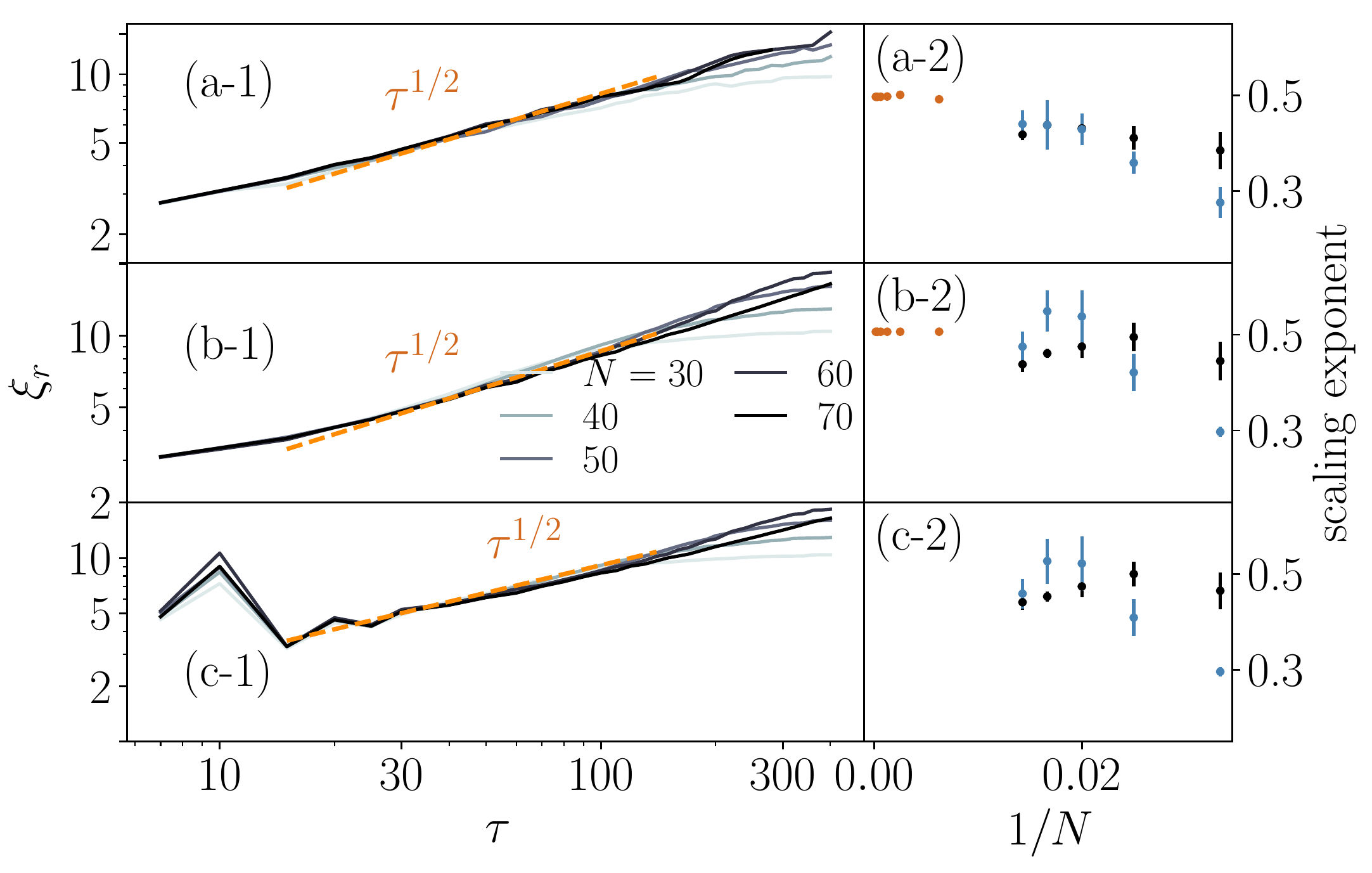}
	\caption{(1) Full lines show time-averaged $\xi_r$ after the end of a quench of (a) the LCM, (b) the real-space density of excitations, and (c) the orbital polarization as a function of $\tau$ for different $N$, plotted in the log-log scale. Average is performed over 20 disorder realizations with $\delta u_0=0.05$. The orange dashed lines represent the $\tau^{1/2}$ scaling. (2) The corresponding scaling exponents, estimated as in Fig.~\ref{fig:domGS} using the $\xi_r$ vs. $\tau$ data in intervals $15\leq \tau\leq140$ (black) and $30\leq\tau\leq 300$ (blue). Orange dots in panels (a-2) and (b-2) are exponents extracted from the $\xi_k$ vs. $\tau$ data and the $n_\textrm{exc}^{-1/2}$ vs. $\tau$ data, respectively, after the end of a quench in the clean system.}
	\label{fig:scaling}
\end{figure}

The post-quench inhomogeneities are larger for quenches performed more slowly (see Supplemental Material S2 \cite{SM}). Their size exhibits a power-law scaling with quench time as shown in Fig.~\ref{fig:scaling}(a-1). The scaling exponent [Fig.~\ref{fig:scaling}(b-1)] lies somewhere between 0.4 and 0.5 for the largest system considered here, depending on the chosen fitting range. In the absence of disorder much larger systems can be considered and the scaling exponent extracted from the $\xi_k$ vs. $\tau$ data (orange dots) is close to $1/2$ (0.497 for $N=6400$). 

	
Inhomogeneities appear also in the density of excitations
$n_\mathrm{exc}(\mathbf{r},t)=\sum_{\sigma}\langle\mathbf{r},\sigma|\hat{P}(t)\hat{P}_\mathrm{c}(t)|\mathbf{r},\sigma\rangle$
where $\hat{P}_c(t)$ is the projector onto the instantaneous conduction band (see Supplemental Material S5 \cite{SM}), and in the deviation of the orbital polarization  
$p(\mathbf{r},t)=\sum_\sigma\langle\mathbf{r},\sigma|\hat P(t)\hat\sigma_z|\mathbf{r},\sigma\rangle$ from the ground state one (see Supplemental Material S6 \cite{SM}). The profiles of the excitations and the polarization appear to be the same as those of the LCM. The scaling of $\xi_r$ of these quantities with quench time also conforms to the KZ prediction, as shown in Figs.~\ref{fig:scaling}(b) and \ref{fig:scaling}(c). 

$\xi_r$ in Fig.~\ref{fig:scaling} saturates at a certain $\tau$. This is a finite-size effect: the value of $\tau$ where this happens increases with the system size $N$.  

The emergence of the KZ length scale is observed also after quenches across $u_c=0$ where topological phases with $C=-1$ and $C=1$ meet (see Supplemental Material S7 \cite{SM}).

\textit{Discussion.}---The scaling of $\xi_r$ can be explained by the Landau-Zener dynamics, taking additionally into account the effect of disorder on eigenstates of the post-quench Hamiltonian.  
Because the disorder is weak, these eigenstates maintain a definite value of the momentum magnitude $|\mathbf k|$  up to a good approximation. The post-quench probability of an excitation is therefore given by the Landau-Zener formula $\exp(-\frac{2|\mathbf k|^2\tau}{|u_1-u_0|})$. This provides a rough estimate for the highest momentum where excitations are present, $k_\mathrm{max}=(\frac{|u_1-u_0|}{2\tau})^{1/2}$. 

In the presence of weak disorder, the eigenstates preserve the length scale $\frac{2\pi}{|\mathbf k|}$. The post-quench projector to the occupied subspace $\hat P(\tau)$ is expected to contain within itself the finest of those length scales present, $\frac{2\pi}{k_\mathrm{max}}\propto\tau^{1/2}$. $\hat P(\tau)$ is an ingredient of expressions for all the quantities studied in this paper, which consequently after a quench exhibit the $\tau^{1/2}$ scaling of $\xi_r$.

\textit{Conclusions.}-- We investigated the equilibrium and the
dynamical critical properties in a Chern insulator as well as their
relation via the KZ mechanism.  We used a weak disorder to reveal the
correlation length scale in the ground-state real-space profile of the
LCM. After a quench the LCM exhibits inhomogeneities of a length scale
that grows approximately as a square root of the quench time. We followed the
growth of the inhomogeneities during the quench and demonstrated that
the KZ freeze-out mechanism applies. Through the lens of the LCM the
critical behavior of weakly disordered Chern insulators is analogous
to the one found in systems with spontaneously broken symmetries. The
important difference is that the amplitude of the inhomogeneities
vanishes with the vanishing disorder strength. Whereas
  proposals for measuring the LCM exist \cite{Caio19}, the KZ scaling
  of inhomogenities with $\tau$ can be seen also in a more directly
  accessible quantities, such as the orbital polarization
  \cite{Sun18} or (if the particle-hole symmetry would be broken, as
  is for instance in the Haldane model) the charge density.
\\
\acknowledgements
We thank R. \v Zitko who provided the initial stimulus for this work. We acknowledge useful discussions with A. Ram\v{s}ak, M. D. Caio, M. J. Pacholski and J. \v{S}untajs. The work was supported by the Slovenian Research Agency under contract no. P1-0044. J. Mravlje acknowledges support by Slovenian Research Agency under Contract No. J1-2458 and L. Ul\v{c}akar support by  L’Oreal-UNESCO For Women in Science Programme.
\\
\\
\begin{center}
\large{\textbf{SUPPLEMENTARY INFORMATION}}
\end{center}
\section{S1.$\quad$ Computation of the LCM on a periodic geometry}

For a system with periodic boundary conditions, the commutators with the position operators $\hat{x}$ and $\hat{y}$ in Eq.~(4) of the main article can be calculated as \cite{Prodan10,Prodan11}               
\begin{equation}
-i[\hat{x},\hat{P}]=\sum_{m=-N/2}^{N/2}c_me^{-im\Delta \hat{x}}\hat{P}e^{im\Delta \hat{x}},
\end{equation}
where $N$ is the number of unit cells in one dimension, $\Delta=2\pi/N$, and $c_m$ are the finite difference coefficients chosen such that $x-\sum_{m=-N/2}^{N/2}c_me^{im\Delta x}=O(\Delta^N)$. Recently, an efficient method was proposed for calculating the average of the LCM of a large section of a system.\cite{Varjas20}

\section{S2.$\quad$ Additional LCM profiles}

\begin{figure}[h]
	\includegraphics[width=0.49\columnwidth]{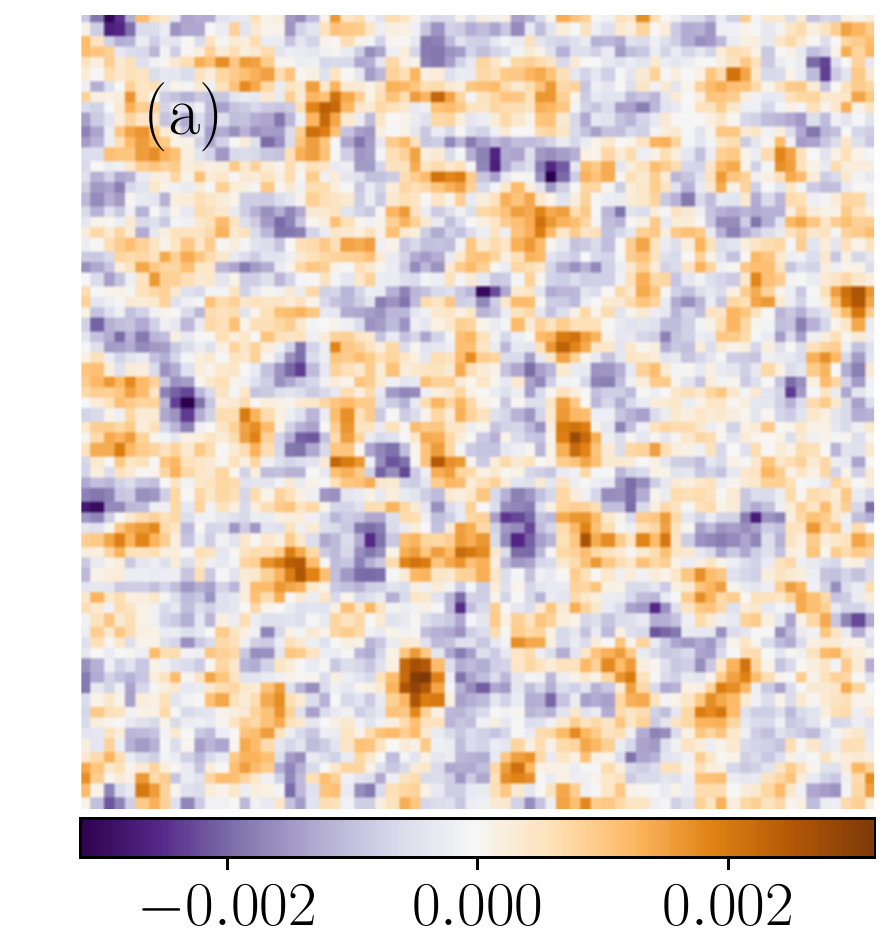}
	\includegraphics[width=0.49\columnwidth]{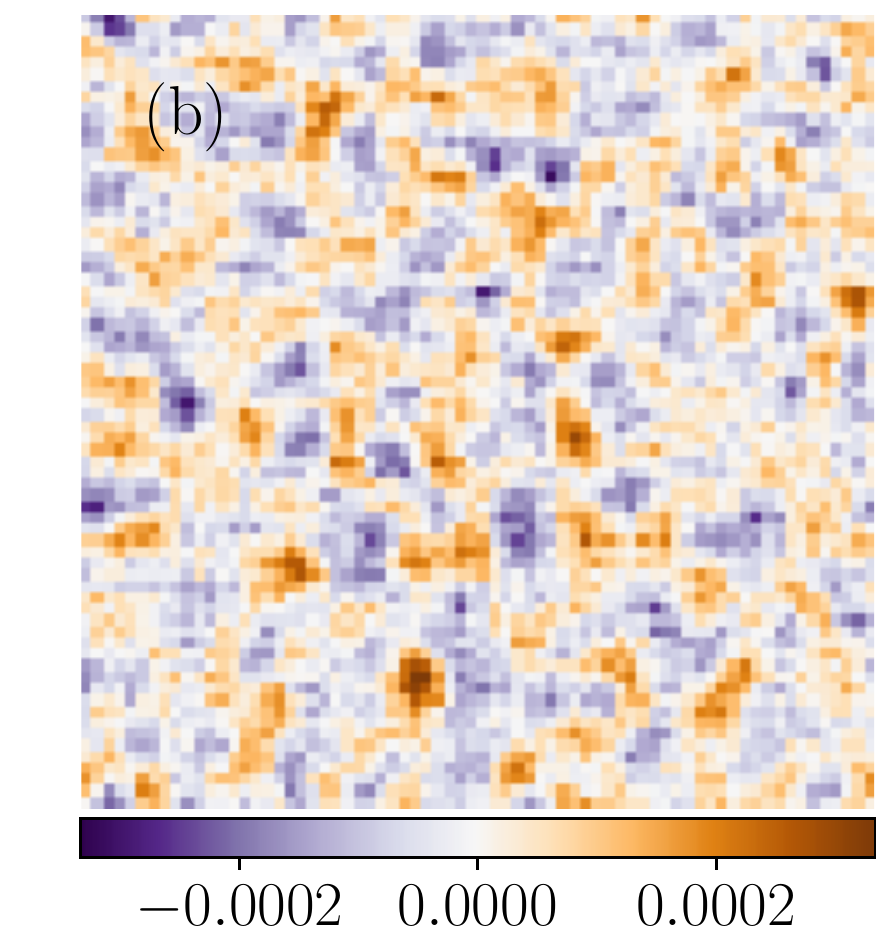}
	\caption{Ground-state LCM profiles at $u=-2.3$ with (a) $\delta u_0=0.05$ and (b) $\delta u_0=0.005$. $N=70$.}
	\label{fig:ampl}
\end{figure}

\begin{figure}[h]
	\includegraphics[width=0.49\columnwidth]{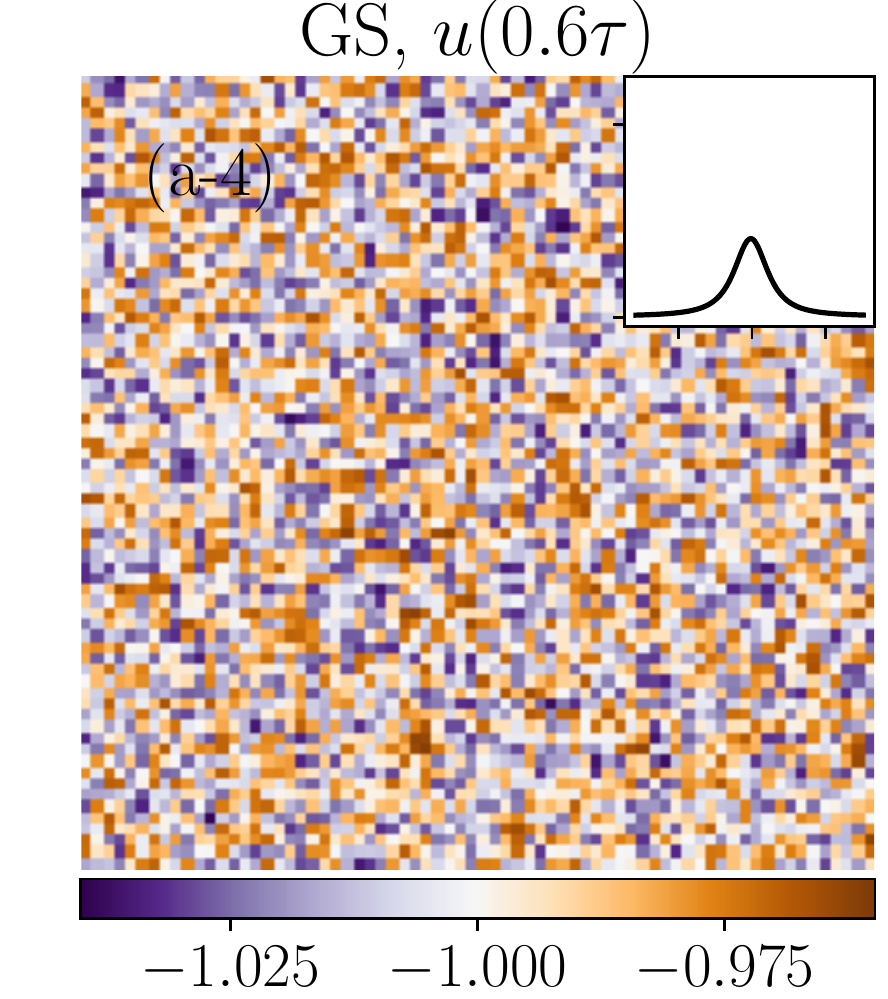}
	\includegraphics[width=0.49\columnwidth]{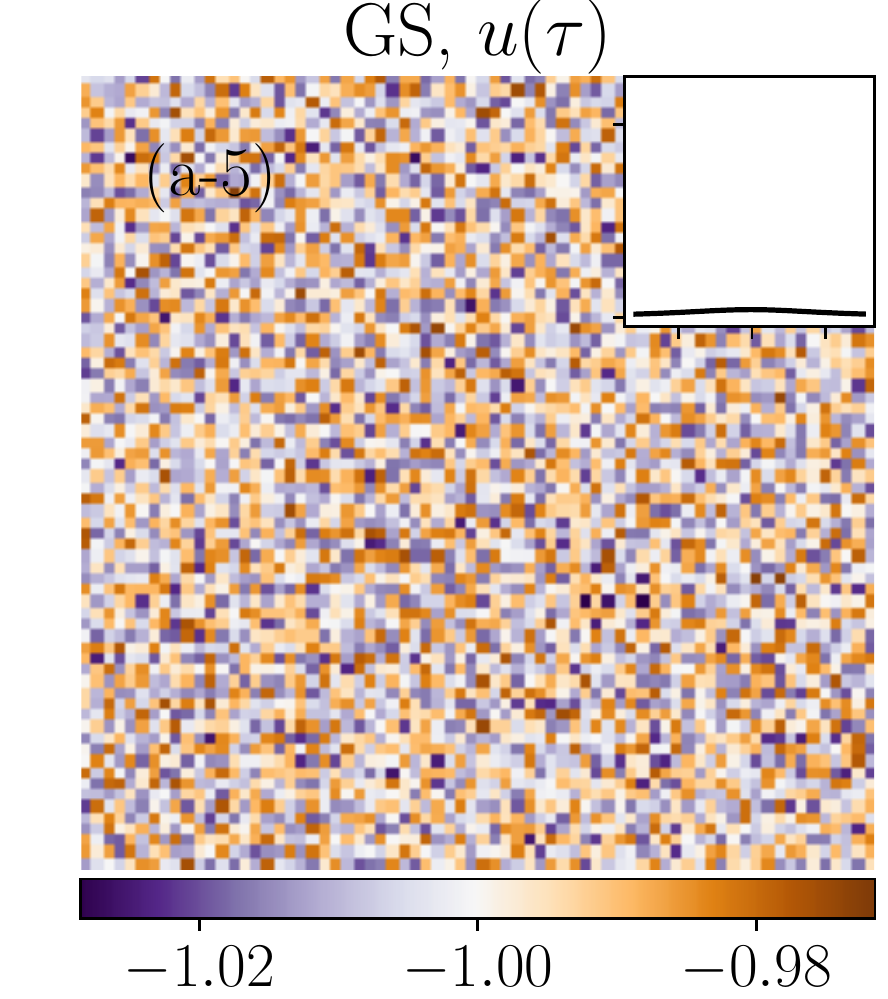}
	\caption{Unfiltered LCM profiles corresponding to Figs.~1(a-4) and 1(a-5) of the main article.}
	\label{fig:subDis}
\end{figure}

\begin{figure}[h]
	\includegraphics[width=0.49\columnwidth]{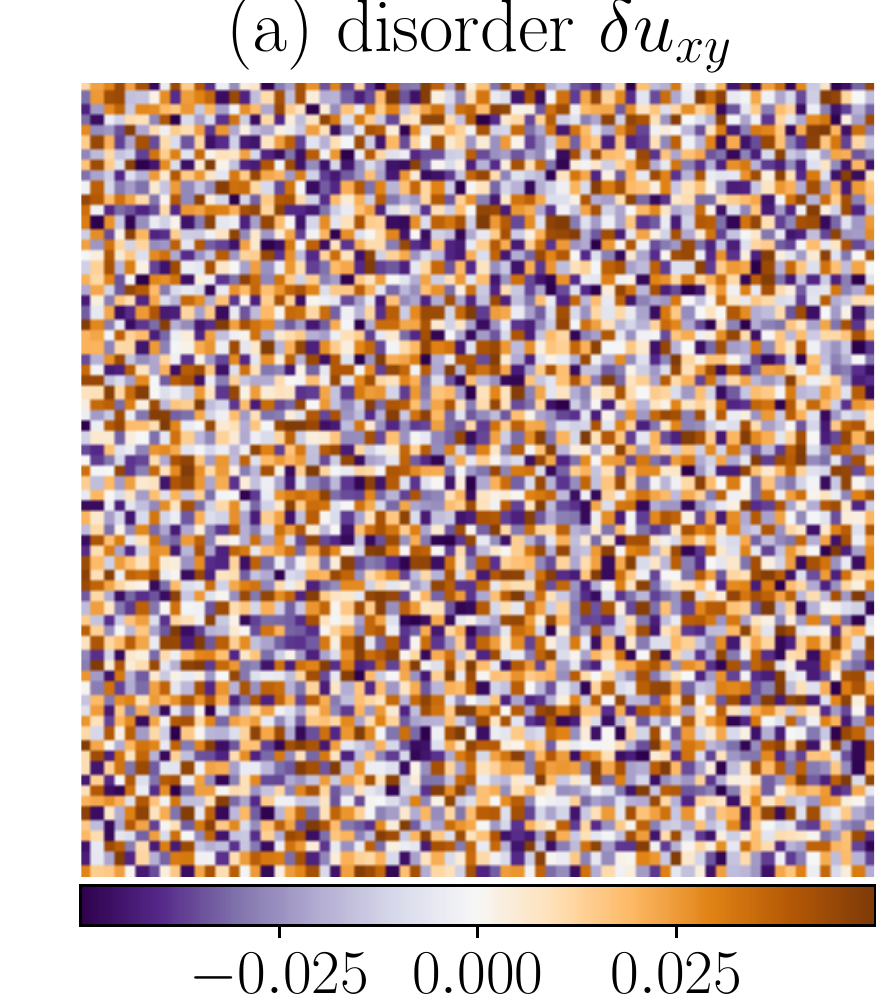}
	\includegraphics[width=0.49\columnwidth]{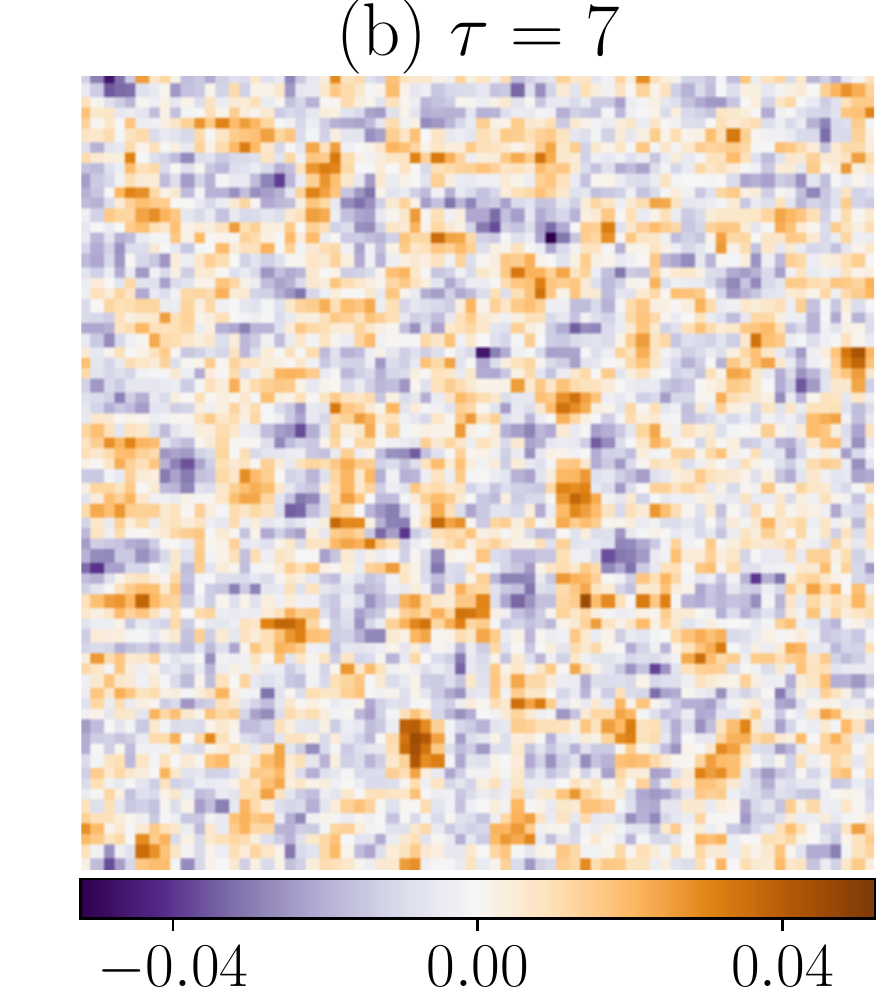}
	\includegraphics[width=0.49\columnwidth]{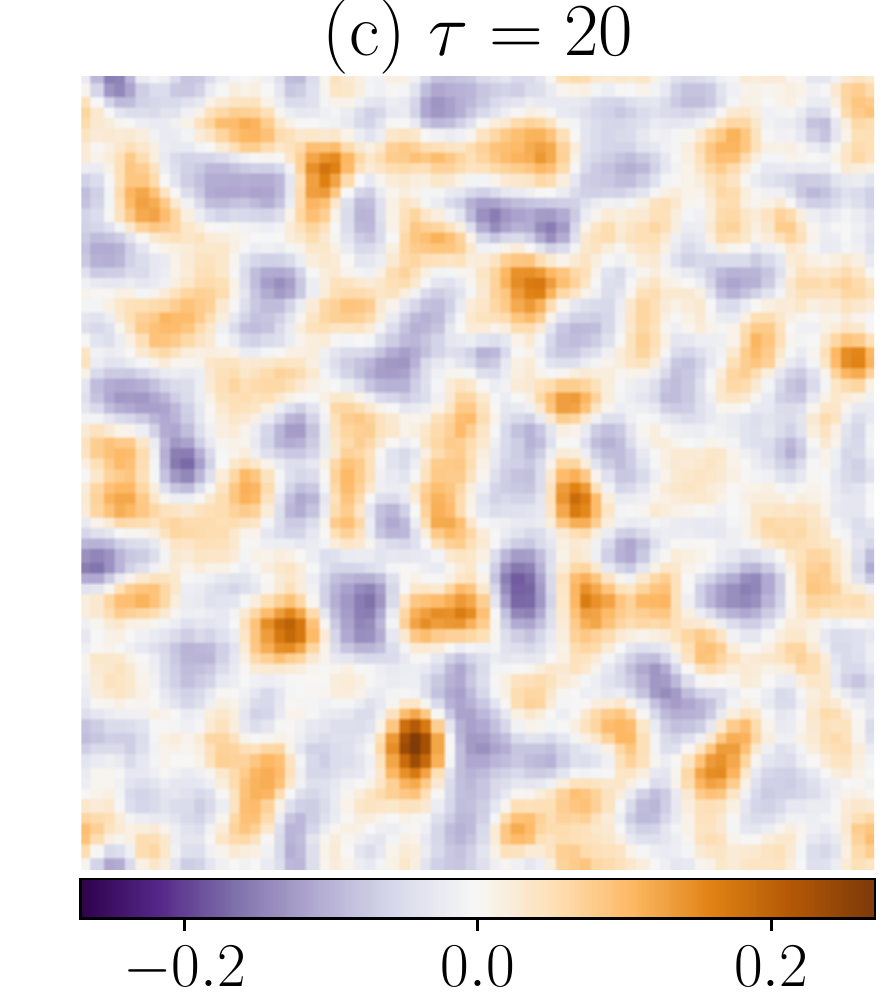}
	\includegraphics[width=0.49\columnwidth]{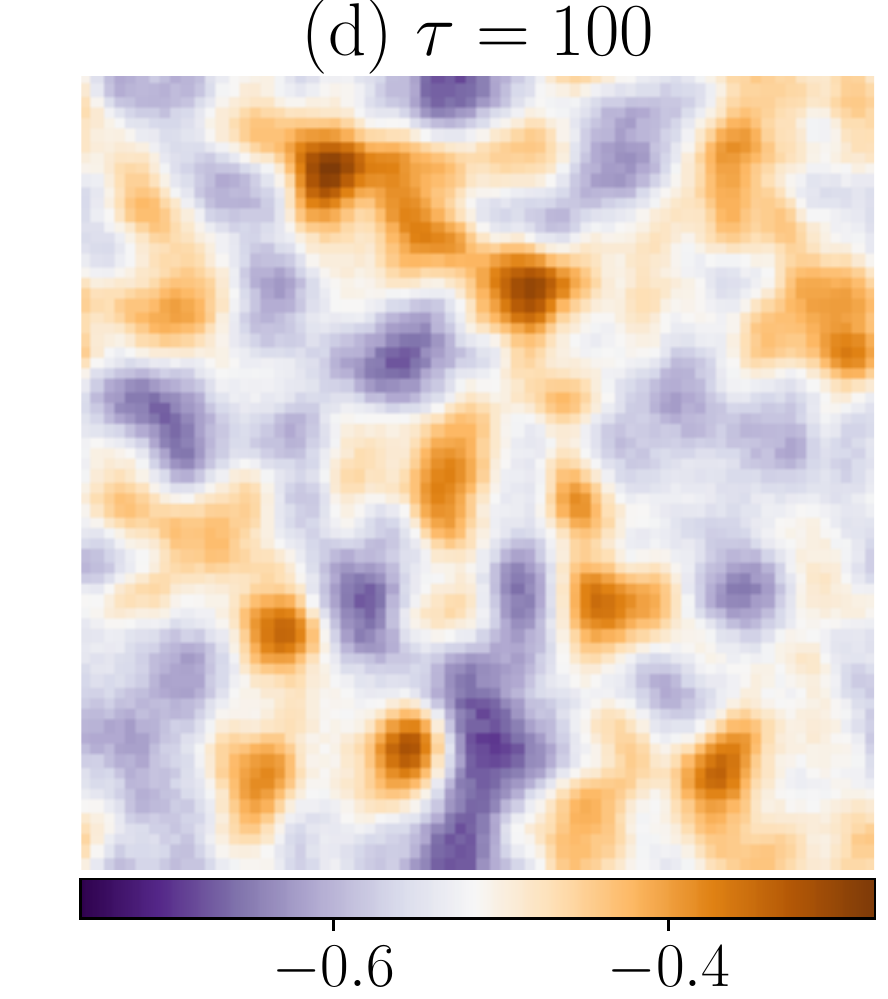}
	\caption{Post-quench LCM profiles for quenches with (b)~$\tau=7$, (c) $\tau=20$ [same as Fig.~1(b-5) of the main article] and (d) $\tau=100$. The disorder realization is shown in panel (a). $N=70$ and $\delta u_0=0.05$.}
	\label{fig:domains}
\end{figure}

Fig.~\ref{fig:ampl} shows that the size of inhomogeneities in the LCM does not depend on the weak disorder strength. On the other hand, the amplitude of deviations is proportional to $\delta u_0$.  
In Fig.~\ref{fig:subDis} the ground-state LCM profiles in the topological phase are shown. They are dominated by a contribution proportional to the disorder. The corresponding profiles in Figs.~1(a-4) and 1(a-5) of the main article were obtained by filtering out this contribution with the Gaussian filter with the width $\sigma=1$ (lattice spacing). For discussion, see Sec. S4.
In Fig.~\ref{fig:domains} post-quench LCM profiles for various quench times are shown for the same disorder realization as in Fig. 1 of the main article. 

\section{S3.$\quad$Estimation of the size of inhomogeneities}
\label{app:size}
\begin{figure}[h]
	\includegraphics[width=0.7\columnwidth]{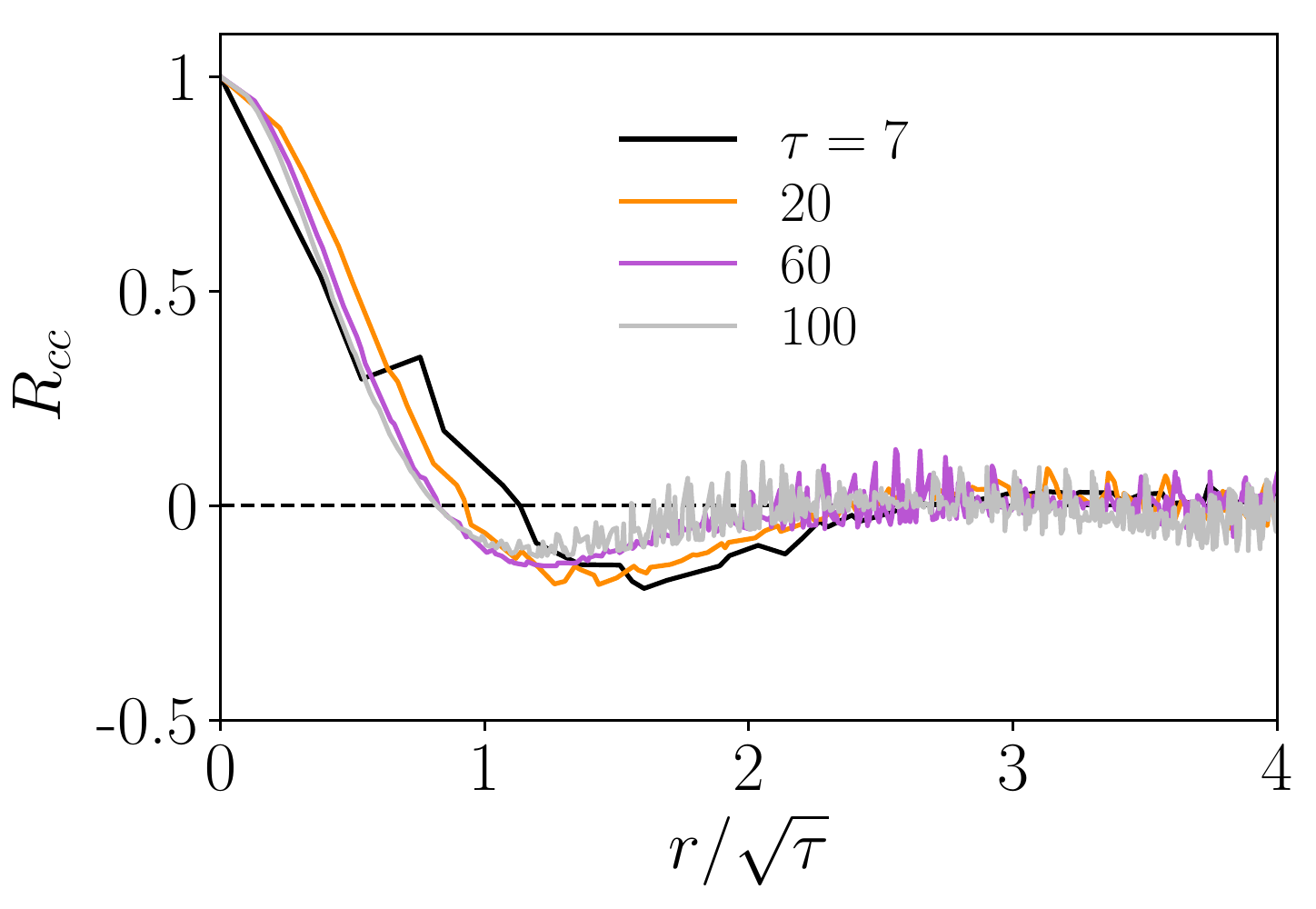}
	\caption{Autocovariance functions of the post-quench LCM profiles for the disorder realization of Fig.~\ref{fig:domains}(a). The profiles for $\tau=7$, 20 and 100 are shown in Figs.~\ref{fig:domains}(b), \ref{fig:domains}(c) and \ref{fig:domains}(d), respectively.}
	\label{fig:cov}
\end{figure}

We estimate the size of the inhomogeneities in a real-space profile $A(\mathbf{r})$ from its position autocovariance function 
\begin{equation}
R_{AA}(r)=\frac{\sum_{|\mathbf{r}|=r}\sum_{\mathbf{r'}}A(\mathbf{r'})A(\mathbf{r}+\mathbf{r'})}{\sum_{|\mathbf{r}|=r}\sum_{\mathbf{r'}}A(\mathbf{r'})^2}.
\end{equation} 
We identify the typical length scale $\xi_r$ in the LCM as the distance at which the disorder-averaged autocovariance function crosses zero, $R_{cc}(\xi_r)=0$. The autocovariance functions of the post-quench LCM profiles shown in Fig.~\ref{fig:domains} are plotted in Fig.~\ref{fig:cov} as a function of $r/\sqrt{\tau}$. Note that for slow enough quenches, the size of inhomogeneities $\xi_r$ scales as $\tau^{1/2}$. 

\section{S4.$\quad$ Ground-state LCM profiles around a single impurity}

Here we study deviations $\delta c_1(\mathbf r)$ of the LCM from its clean system value, induced by a single weak impurity with strength $\delta u$ at $\mathbf r=0$. Only the ground state is considered.

\begin{figure}[h]
	\includegraphics[width=0.425\columnwidth]{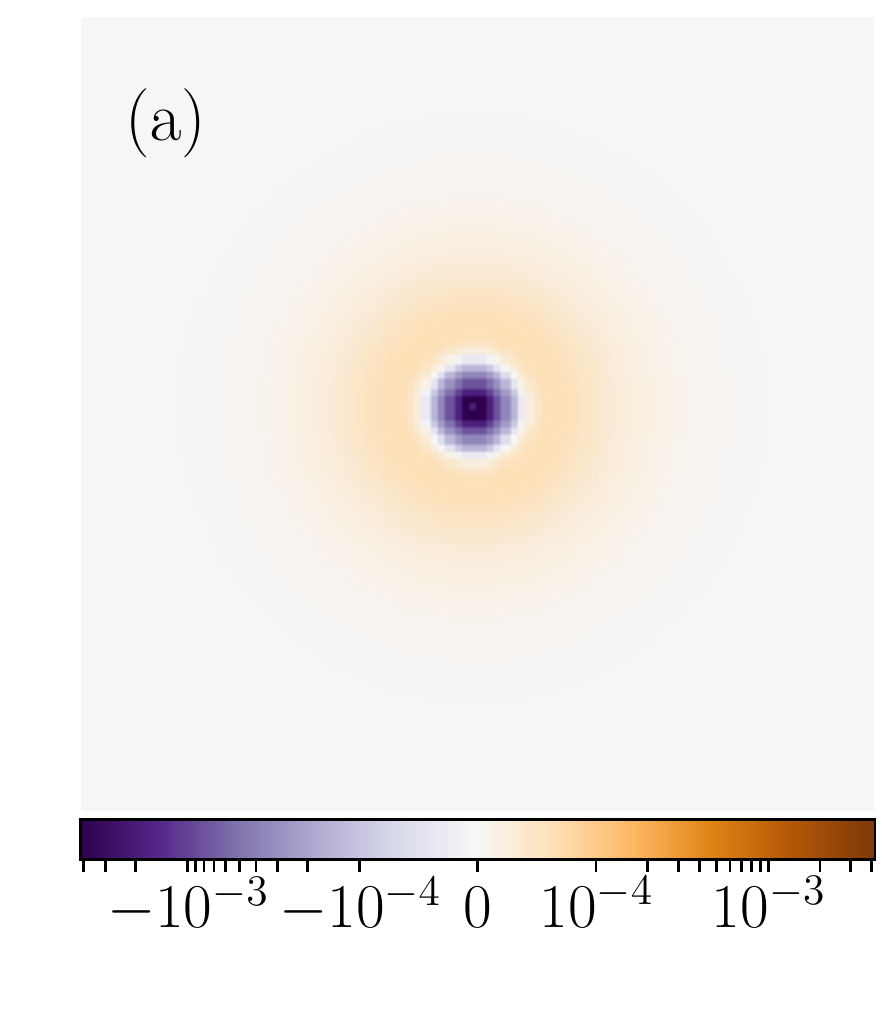}
	\includegraphics[width=0.555\columnwidth]{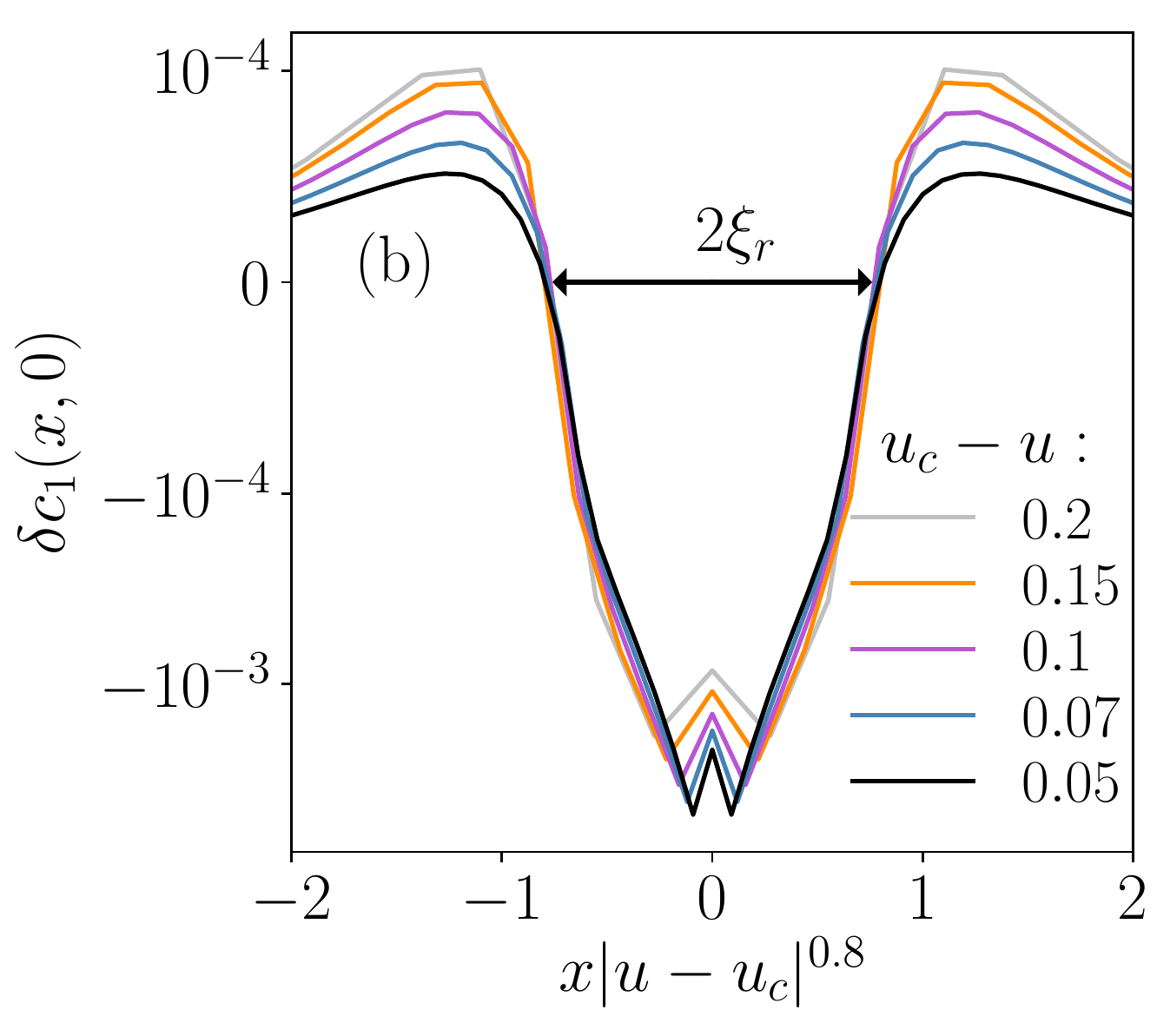}
	\includegraphics[width=0.67\columnwidth]{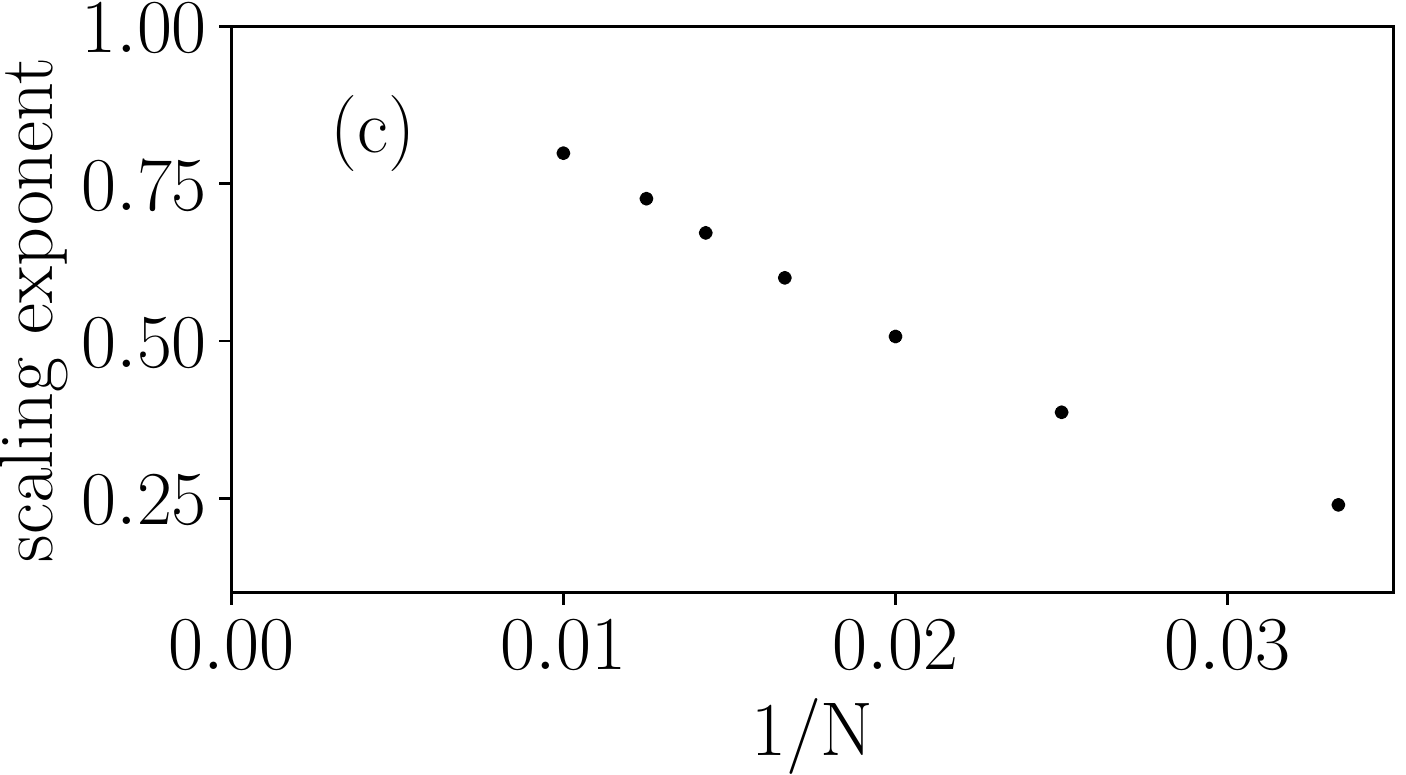}
	\caption{(a) $\delta c_1(\mathbf r)$ around an impurity with strength $\delta u=0.1$ at $\mathbf r=0$. $N=100$ and $u=-2.07$. (b) $\delta c_1(\mathbf r)$ around the same impurity plotted along the $y=0$ line as a function of $x|u-u_c|^{0.8}$ for various $u$ in the trivial phase. $N=100$. In both (a) and (b), linear scale is used for $|\delta c_1(\mathbf r)|<10^{-4}$ and log scale elsewhere. (c) The scaling exponent, estimated by fitting a power-law to the $\xi_r$ vs. $|u-u_c|$ data in the interval $0.02\leq |u-u_c|\leq0.1$ and plotted as a function of $1/N$.} 
	\label{fig:cPoint}
\end{figure}

\begin{figure}[h]
	\includegraphics[width=0.425\columnwidth]{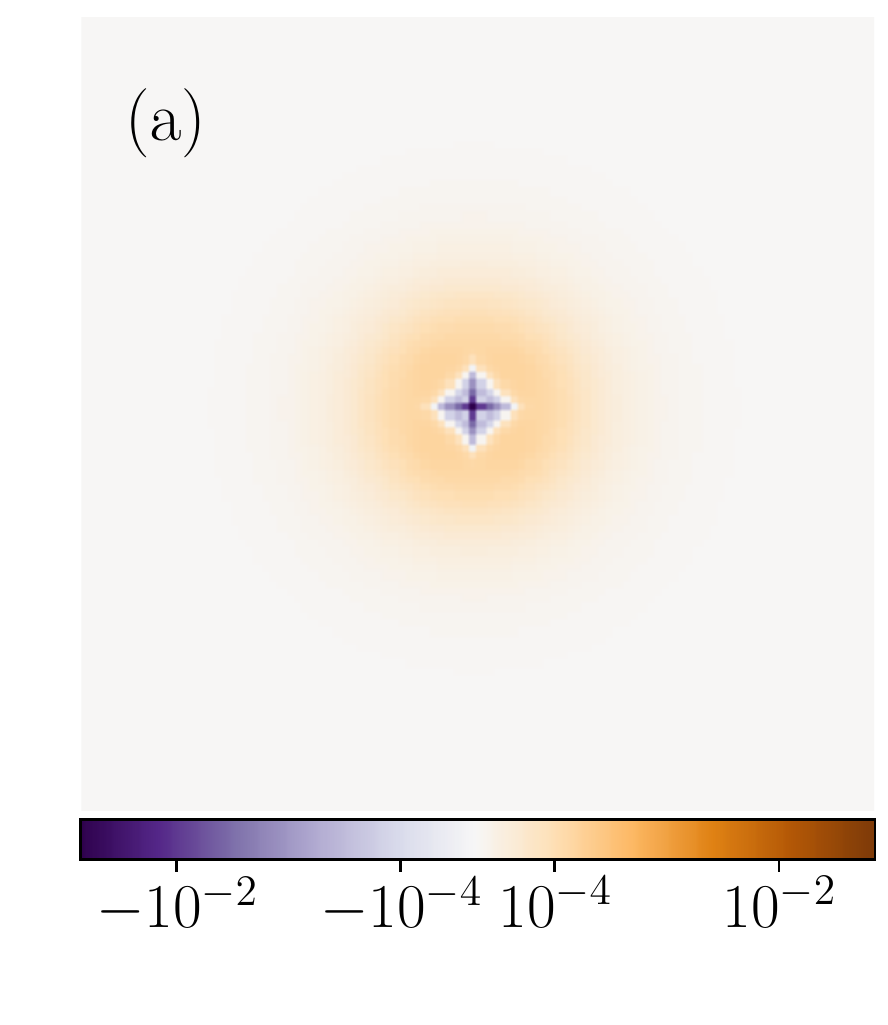}
	\includegraphics[width=0.555\columnwidth]{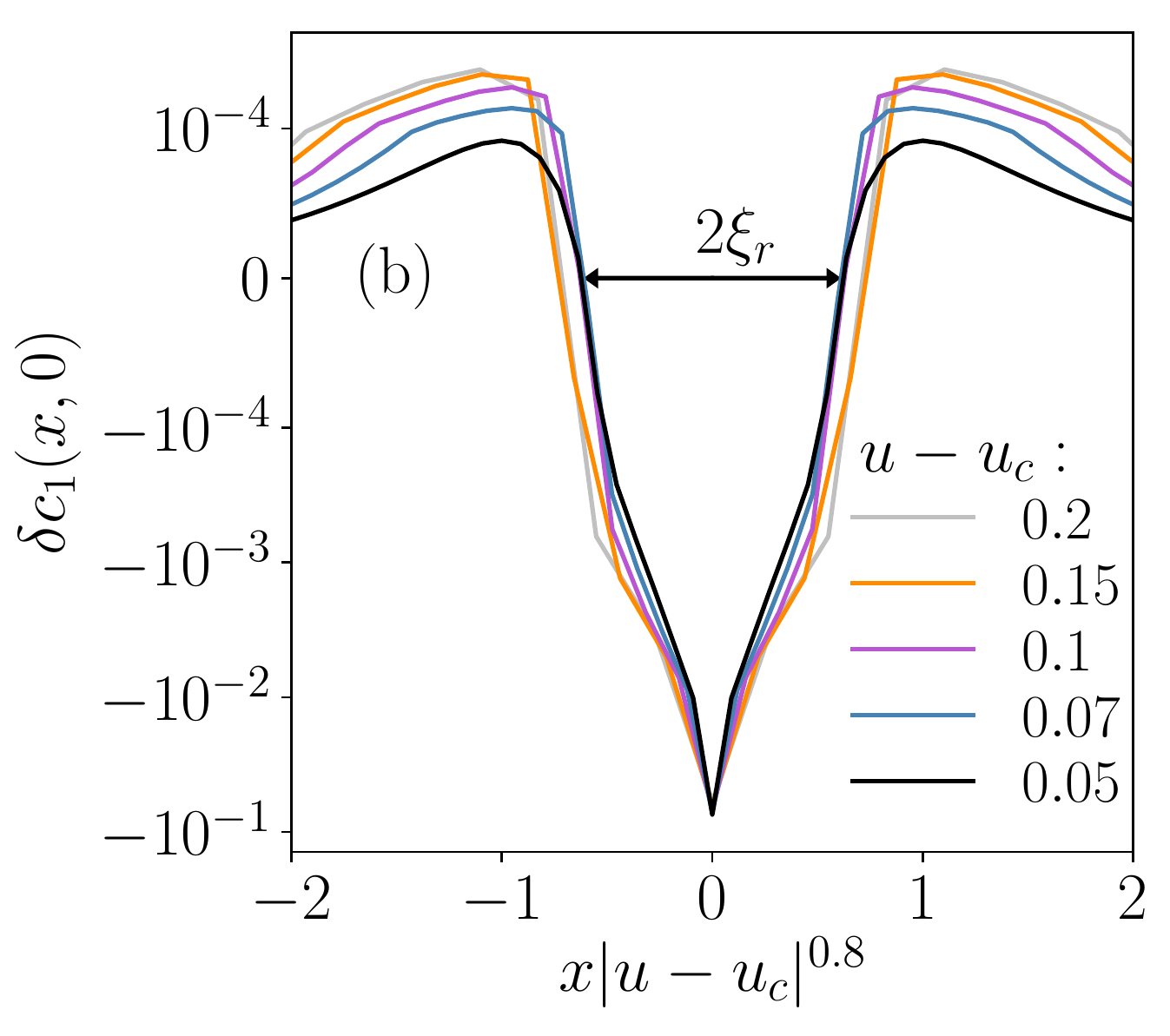}
	\caption{(a) $\delta c_1(\mathbf r)$ around an impurity with strength $\delta u=0.1$ at $\mathbf r=0$. $N=100$ and $u=-1.93$. (b) $\delta c_1(\mathbf r)$ around the same impurity plotted as a function of $x|u-u_c|^{0.8}$ along the $y=0$ line for various $u$ in the topological phase. $N=100$. In both (a) and (b), linear scale is used for $|\delta c_1(\mathbf r)|<10^{-4}$ and log scale elsewhere.}
	\label{fig:cPoint1}
\end{figure}

In Figs.~\ref{fig:cPoint}(a) and \ref{fig:cPoint1}(a), typical $\delta c_1(\mathbf r)$ profiles in the trivial and in the topological phases, respectively, are shown. In Figs.~\ref{fig:cPoint}(b) and \ref{fig:cPoint1}(b), $\delta c_1(\mathbf r)$ are plotted along the $y=0$ line for several values of $u$ in the trivial and in the topological phases, respectively. The position is rescaled as $x|u-u_c|^{0.8}$, showing that the radius $\xi_r$ of the region around the impurity where the LCM deviates below (for $\delta u>0$) the clean system value scales as a power law. The estimate of the scaling exponent on the trivial side approaches one as the system size is increased, see Fig.~\ref{fig:cPoint}(c). Note, however, that $\xi_r$ is not the only scale in which $\delta c_1(\mathbf r)$ can be described: it has also an internal structure.   

A notable feature is that the behavior of $\delta c_1(\mathbf r)$ on the topological and on the trivial side are quite different. On the topological side, $|\delta c_1(\mathbf r)|$ is maximal at the position of the impurity. There it takes a value that is at least five times larger than the value on neighboring sites, whereas on the trivial side it has a minimum at the position of the impurity and takes a maximum on the neighboring sites. This distinction has important consequences for the behavior of the LCM in a disordered system. Namely, for a weak disorder one can write $\delta c(\mathbf r) \sim \int\mathrm d\mathbf r' \delta c_1(\mathbf r-\mathbf r')\delta u(\mathbf r')$ where $\delta c(\mathbf r)$ is the deviation of the LCM in the disordered system from the clean system value and $\delta u(\mathbf r)$ the distribution of the disorder. $\delta c_1(\mathbf r)$ thus plays the role of the integration kernel through which the disorder is averaged. Now, on the topological side, because the largest contribution to $\delta c_1(\mathbf r)$ is local, one can expect that the $\delta c(\mathbf r)$ is dominated by a contribution directly proportional to disorder. This is indeed what one observes in Fig.~S2. 

\section{S5.$\quad$ Real-space distribution of excitations}

In this section we derive the real-space distribution of the excitations in the conduction band. Let us first consider a clean system. Let $|\Psi_\alpha(\mathbf{k},t)\rangle=|\mathbf{k¸}\rangle\otimes|\psi_\alpha(\mathbf{k},t)\rangle$ be the instantaneous eigenstates, i.e., the eigenstates of the Hamiltonian $H(t)$, and $|\Phi_\alpha(\mathbf{k},t)\rangle=|\mathbf{k}\rangle\otimes|\varphi_\alpha(\mathbf{k},t)\rangle$ the states obtained by time-evolving the pre-quench eigenstates $|\Psi_\alpha(\mathbf k,0)\rangle$ to time $t$. Here $|\mathbf k\rangle=\frac{1}{N^2}\sum_\mathbf{r}e^{i \mathbf k\cdot \mathbf r}|\mathbf r\rangle$ and $\alpha=v$ and $\alpha=c$ denote the valence and the conduction bands, respectively. The total number of excitations is
\begin{equation}
\begin{split}
N_\mathrm{exc}(t)&=\sum_{\mathbf{k}}|\langle \Psi_c(\mathbf{k},t)|\Phi_v(\mathbf{k},t)\rangle|^2=\\
&=\sum_{\mathbf{k},\mathbf{k}'}\langle \Psi_c(\mathbf{k},t)|\Phi_v(\mathbf{k}',t)\rangle\langle \Phi_v(\mathbf{k}',t)|\Psi_c(\mathbf{k},t)\rangle,\label{eq:Nexc1}
\end{split}
\end{equation} 
where the second line owes to the orthogonality of the plane waves. Recognizing $\hat{P}(t)=\sum_{\mathbf{k}}|\Phi_v(\mathbf{k},t)\rangle\langle\Phi_v(\mathbf{k},t)|$ as the projector onto the occupied subspace and $\hat{P}_\mathrm{c}(t)=\sum_{\mathbf{k}}|\Psi_c(\mathbf{k},t)\rangle\langle\Psi_c(\mathbf{k},t)|$ as the projector onto the instantaneous conduction band, Eq.~\eqref{eq:Nexc1} may be written as a trace over the whole Hilbert space:
$N_\mathrm{exc}(t)=\mathrm{Tr}\{\hat{P}(t)\hat{P}_\mathrm{c}(t)\}$.
We calculate the trace in the real-space basis, 
$N_\mathrm{exc}(t)=\sum_{\mathbf{r},\sigma}\langle\mathbf{r},\sigma|\hat{P}(t)\hat{P}_\mathrm{c}(t)|\mathbf{r},\sigma\rangle$. The real-space distribution of excitation is thus 
\begin{equation}
n_\mathrm{exc}(\mathbf{r},t)=\sum_\sigma\langle\mathbf{r},\sigma|\hat{P}(t)\hat{P}_\mathrm{c}(t)|\mathbf{r},\sigma\rangle. 
\end{equation}
This expression can also be evaluated in a disordered system using $\hat P(t)=\sum_{n\in v}|\Phi_n(t)\rangle\langle\Phi_n(t)|$ and $P_{\mathrm c}(t)=\sum_{n\in c}|\Psi_n(t)\rangle\langle\Psi_n(t)|$ where $|\Psi_n(t)\rangle$ and $|\Phi_n(t)\rangle$ are instantaneous eigenstates and time-evolved pre-quench eigenstates, respectively.

\begin{figure}[h]
	\includegraphics[width=0.49\columnwidth]{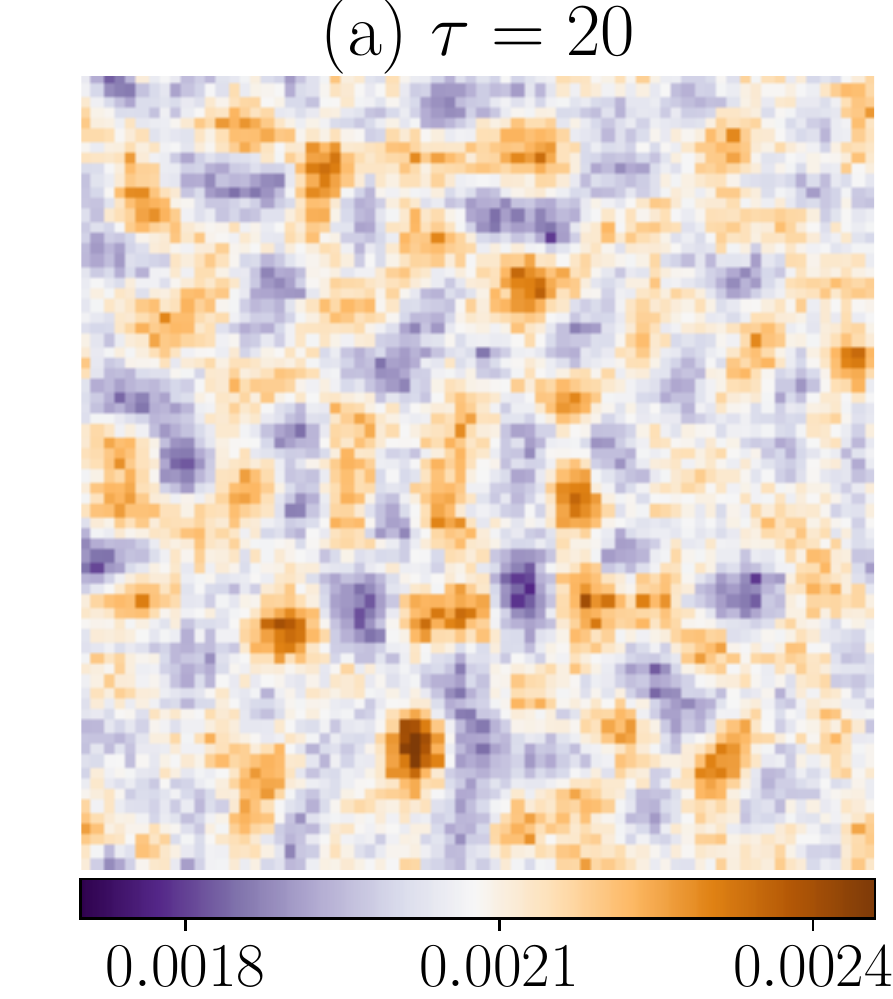}
	\includegraphics[width=0.49\columnwidth]{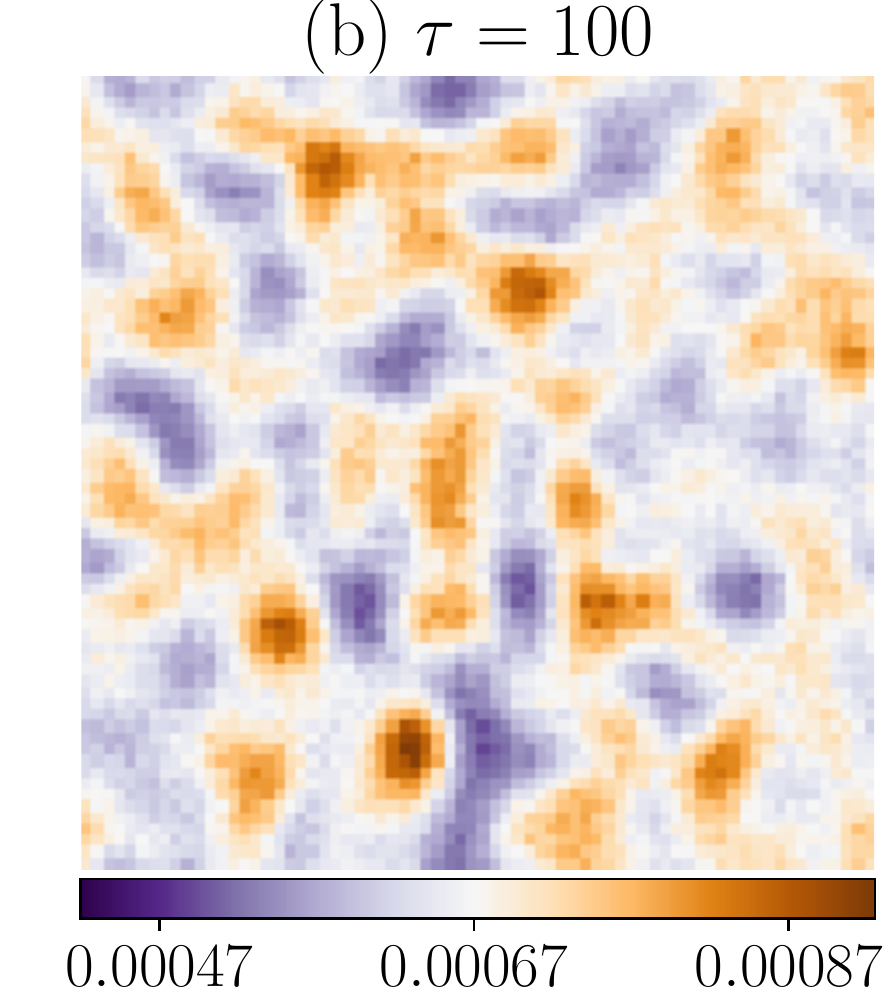}
	\caption{Spatial distribution of excitations after quenches with (a) $\tau=20$ and (b) $\tau=100$. The disorder realization is the same as in Fig.~\ref{fig:domains}.}
	\label{fig:exc}
\end{figure}

In Fig.~\ref{fig:exc} some post-quench profiles of excitations are shown.
\vfill

\section{S6.$\quad$ Real-space distribution of the orbital polarization}

\begin{figure}[h]
	\includegraphics[width=0.49\columnwidth]{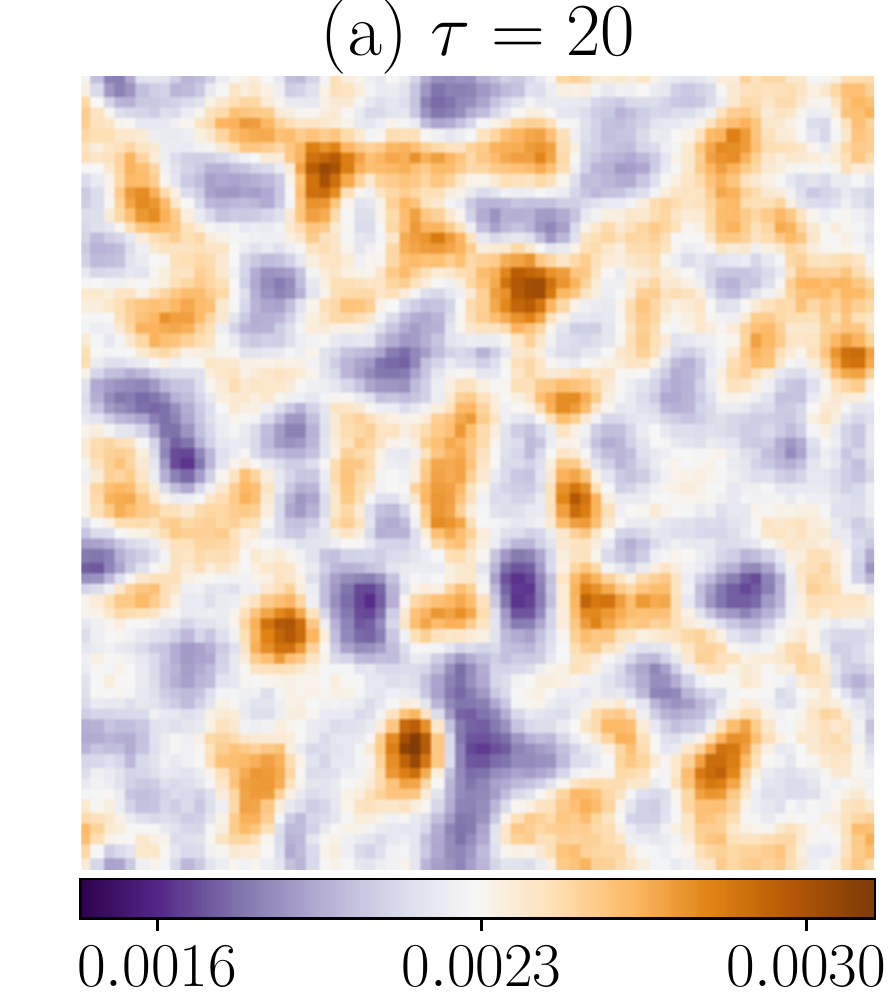}
	\includegraphics[width=0.49\columnwidth]{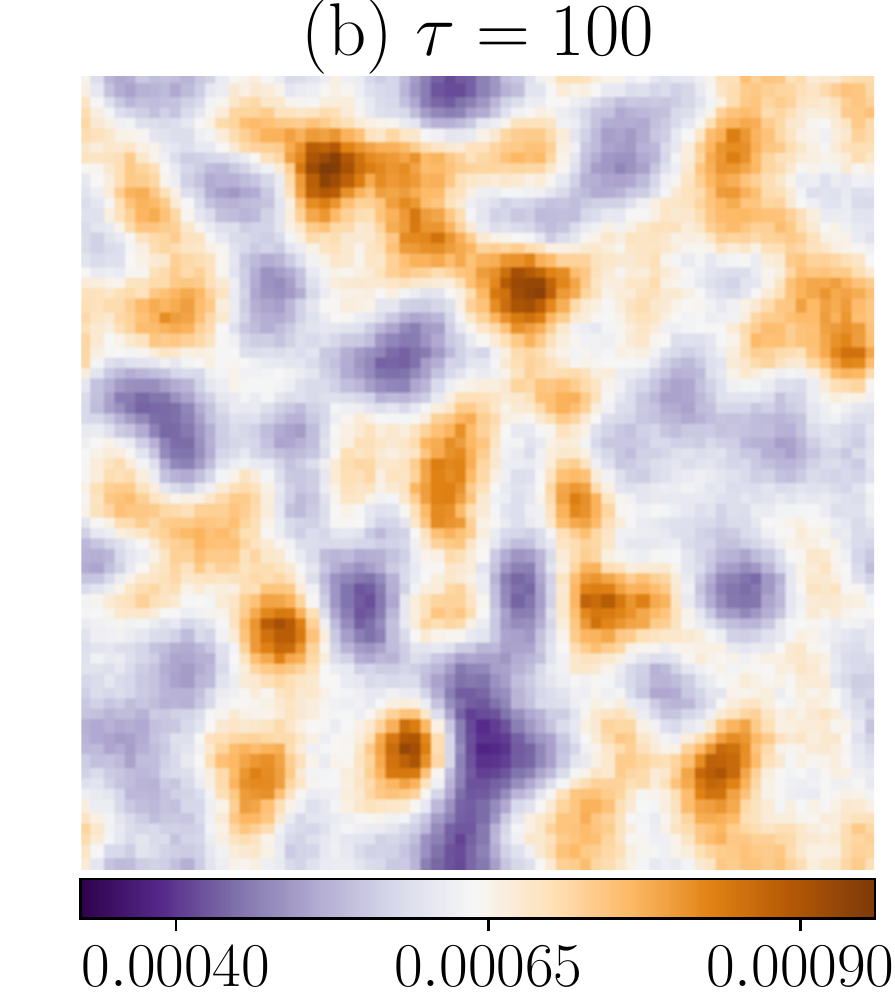}
	\caption{Deviation of the post-quench orbital polarization from the ground-state one for quenches with (a) $\tau=20$ and (b) $\tau=100$. The disorder realization is the same as in Fig.~\ref{fig:domains}.}
	\label{fig:pol}
\end{figure}

We define the orbital polarization as the difference between the occupation of the orbital $A$ and the occupation of the orbital $B$:
\begin{equation}
\begin{split}
p(\mathbf{r},t)&=\sum_{n\in v}|\langle \mathbf{r},A|\Phi_n(t)\rangle|^2-|\langle \mathbf{r},B|\Phi_n(t)\rangle|^2=\\
&=\sum_\sigma\langle\mathbf r,\sigma|\hat P(t)\hat\sigma_z|\mathbf r,\sigma\rangle.
\end{split}
\end{equation}
Some post-quench profiles of the orbital polarization are shown in Fig.~\ref{fig:pol}.

\section{S7.$\quad$ Quench between topological phases}

\begin{figure}[h]
	\includegraphics[width=0.49\columnwidth]{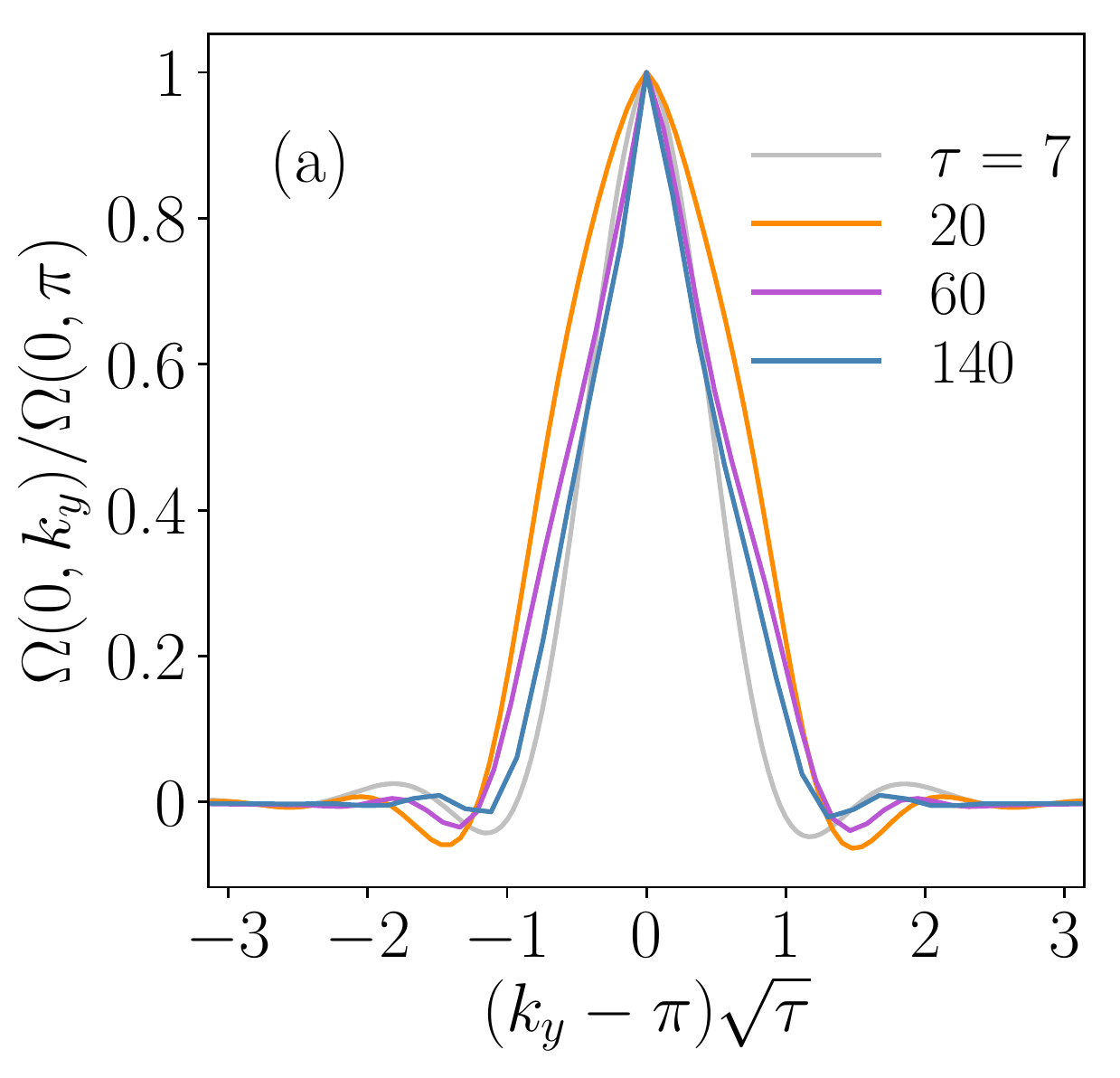}
	\includegraphics[width=0.49\columnwidth]{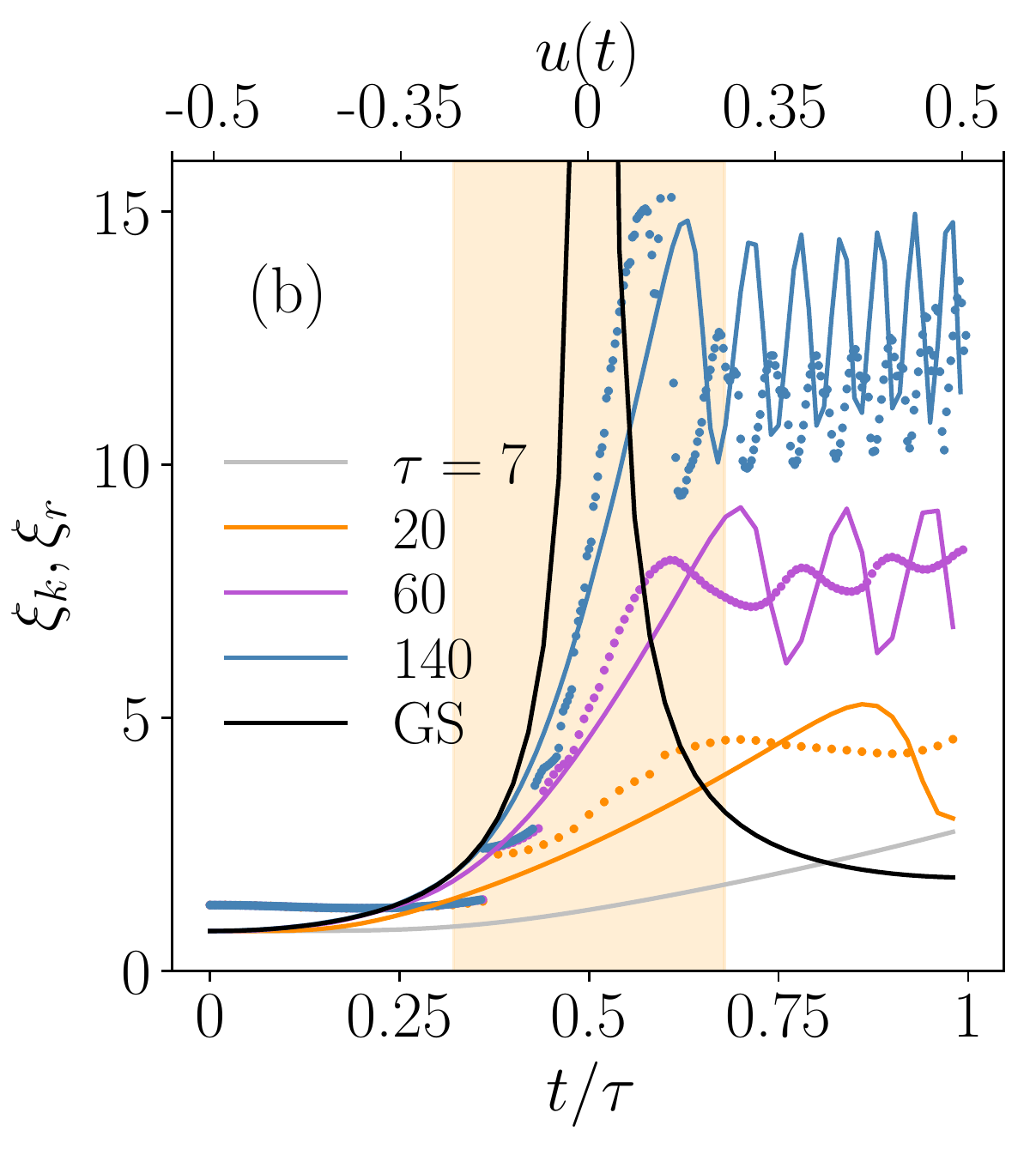}
	\caption{(a) Berry curvature along the $k_x=0$ line at the end of a quench for various $\tau$, plotted as a function of scaled momentum $(k_y-\pi)\sqrt{\tau}$. (b) Full lines show $\xi_k$ during quenches with various $\tau$ (colored) and $\xi_k$ of instantaneous ground states (black). Colored dots show $\xi_r$ of the LCM of a disordered system with $\delta u_0=0.05$ during quenches ($N=70$). The shaded region is the freeze-out zone for the $\tau=20$ quench.}
	\label{fig:Berry1}
\end{figure}
\begin{figure}[h]
	\includegraphics[width=0.49\columnwidth]{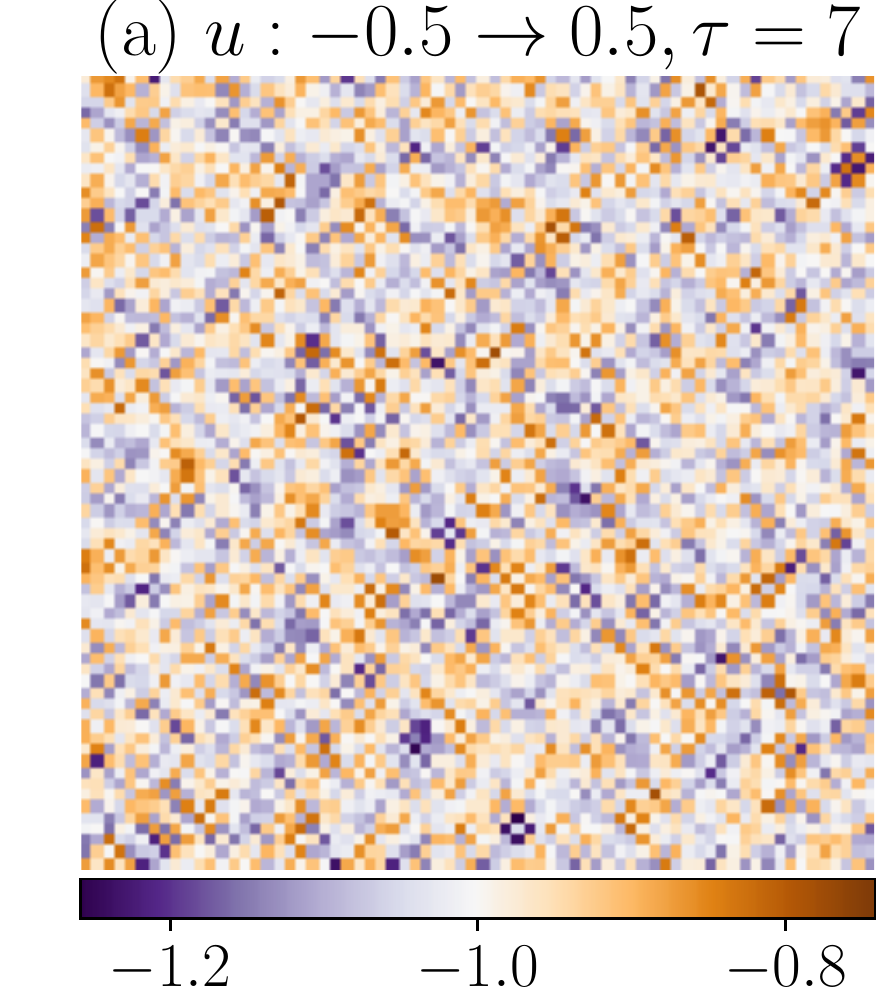}
	\includegraphics[width=0.49\columnwidth]{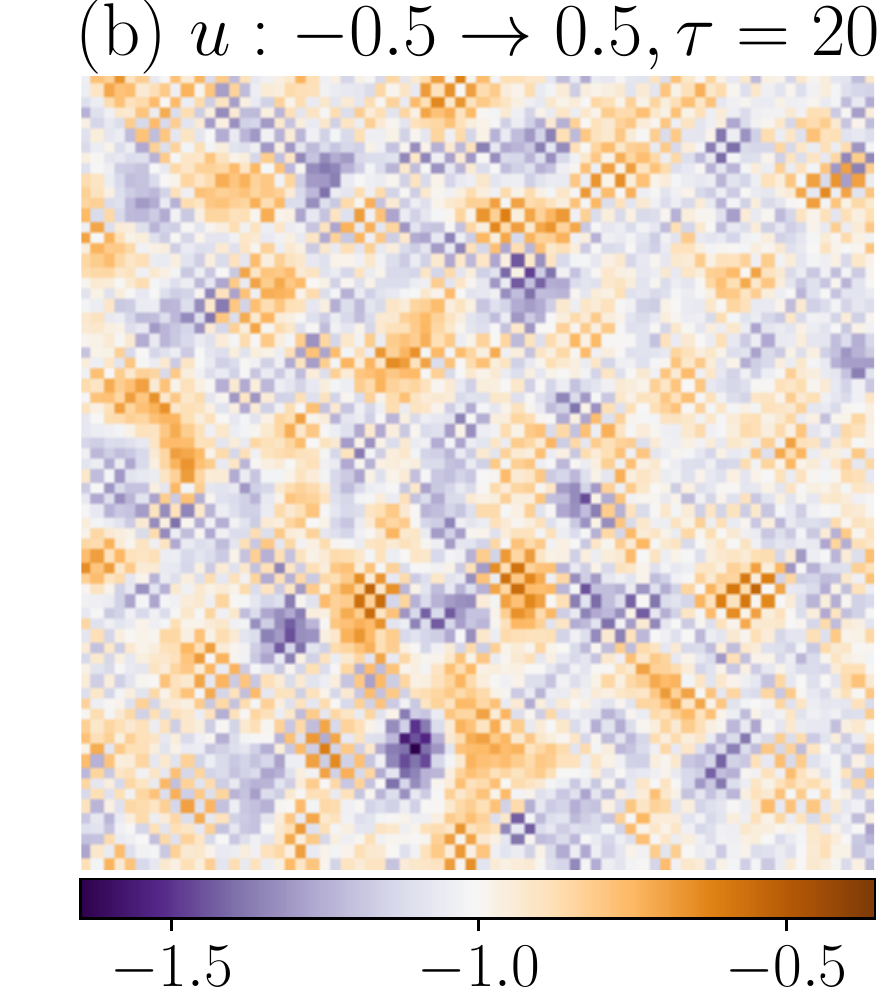}
	\includegraphics[width=0.49\columnwidth]{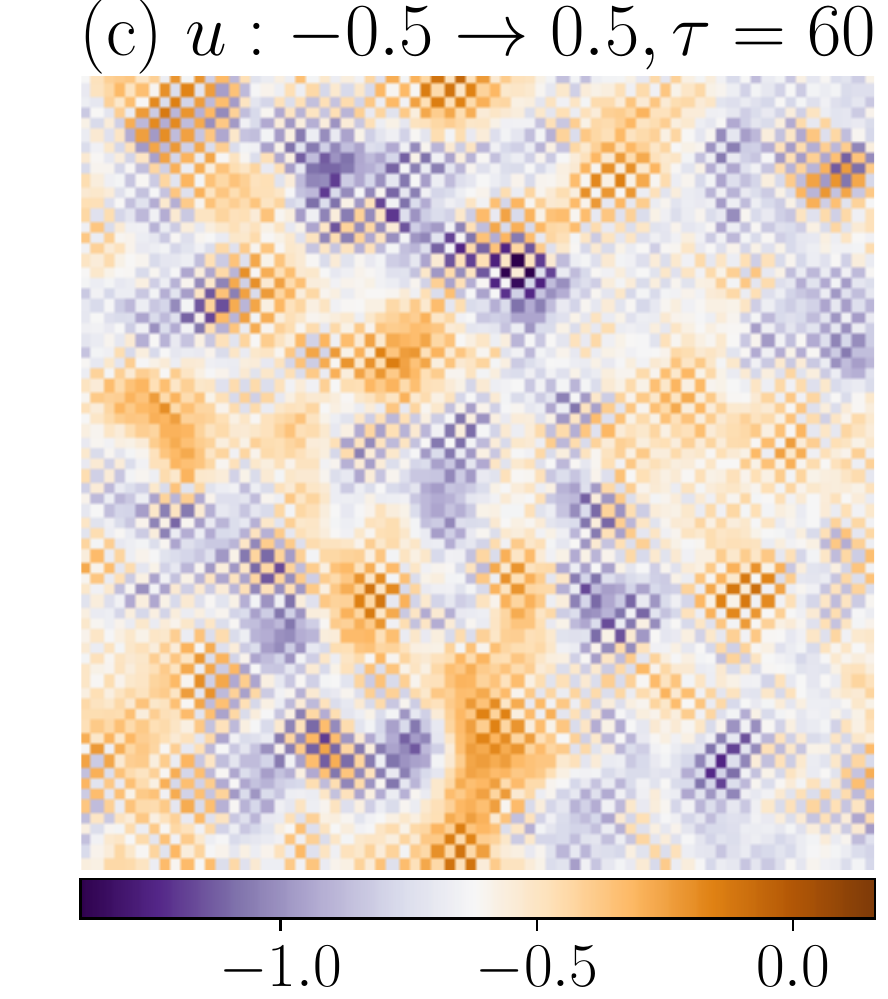}
	\includegraphics[width=0.49\columnwidth]{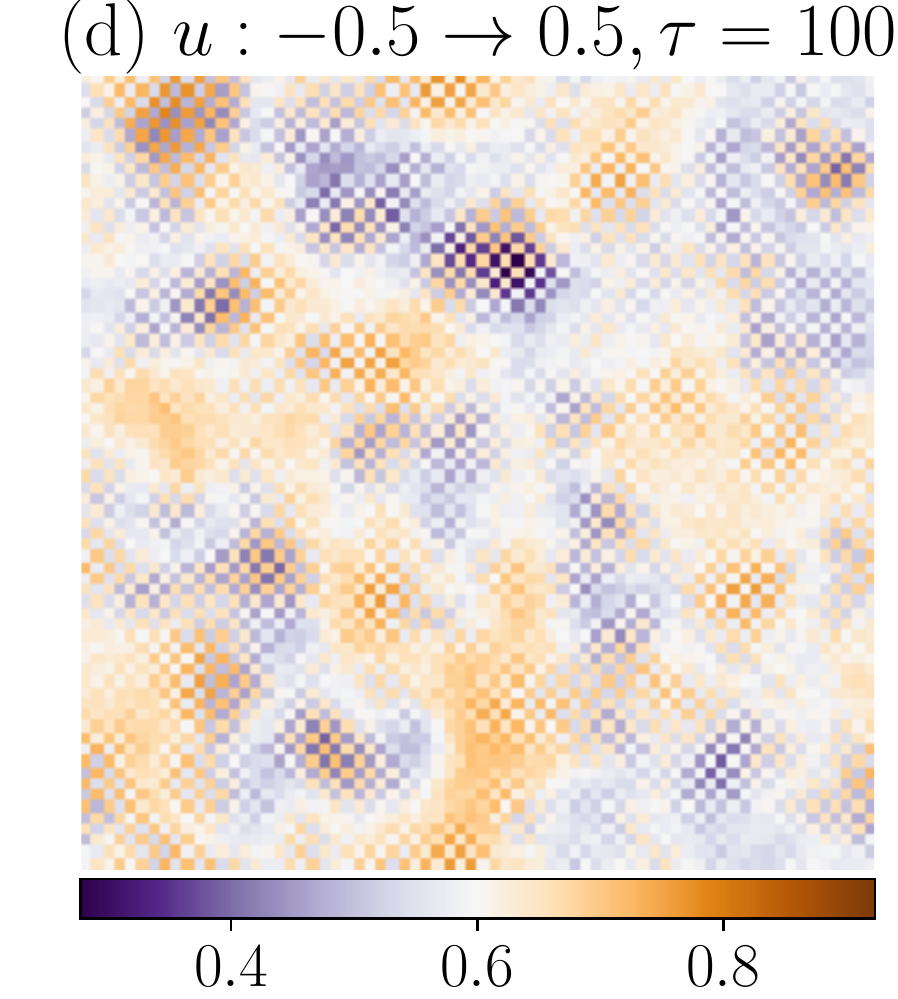}
	\caption{Post-quench LCM profiles for quenches starting at $u_0=-0.5$ and ending at $u_1=0.5$ with (a)~$\tau=7$, (b) $\tau=20$, (c) $\tau=60$ and (d) $\tau=100$. The disorder realization is shown in Fig.~\ref{fig:domains}(a). $N=70$ and $\delta u_0=0.05$.}
	\label{fig:domains1}
\end{figure}
Quenches across a phase transition between two non-trivial phases lead to the same behavior both in clean systems as well as in weakly disordered systems. We investigated a quench starting at $u_0=-0.5$ and ending at $u_1=0.5$. The critical point at $u_c=0$ separates the phases with $C=-1$ ($-2<u<0$) and $C=1$ ($0<u<2$). The energy dispersion forms two Dirac cones at $u_c=0$, centered at $\mathbf{k}=(0,\pi)$ and $\mathbf{k}=(\pi,0)$. The presence of two Dirac points does not modify the critical behavior and the results are thus equivalent to the ones discussed in the main text. The Berry curvature develops peaks about momenta where the energy gap closes. Its value along $k_x=0$, rescaled with $\tau^{1/2}$, is shown in Fig.~\ref{fig:Berry1}(a). At the end of the quench, the width of the peak scales as $\tau^{-1/2}$ and thus $\xi_k\propto\tau^{1/2}$. The time evolution of $\xi_k$ as well as that of $\xi_r$ during the quench is the same as in the case of transitions from trivial to topological phase. These results are presented in Fig.~\ref{fig:Berry1}(b).

Fig.~\ref{fig:domains1} shows the LCM profiles after quenches for various quench times. The disorder realization is the same as in Fig.~1 of the main article. The LCM exhibits inhomogeneities, which become larger for longer quenches. They are modulated by a checkerboard pattern, the emergence of which can be understood because of the presence of two Dirac points at $\mathbf{k}=(0,\pi)$ and $\mathbf{k}=(\pi,0)$ which introduces the corresponding modulation wavevector in the results for weakly disordered system where the translational symmetry is slightly broken. 

\bibliography{bibliography_list}{}

\begin{thebibliography}{75}
\bibitem [{nun()}]{nune1}%
  \BibitemOpen
  \href@noop {} {}\bibinfo {note} {A different critical behavior of the
  curvature function with $\nu\neq 1$ is expected in Chern insulators with
  higher order band crossings \cite{Chen19}.}\BibitemShut {Stop}%
\end{thebibliography}

\end{document}